\documentclass[12pt]{iopart}

\usepackage{amsmath}

\usepackage{iopams}
\usepackage{indentfirst}

\usepackage{bm}			    
\usepackage{bbm}		    
\usepackage{graphicx}

\usepackage{color}
\definecolor{myblue}{RGB}{65,105,225}
\definecolor{mygreen}{RGB}{34,139,34}
\definecolor{myorange}{RGB}{255,69,0}
\usepackage[colorlinks,linkcolor=myorange,urlcolor=myblue,citecolor=myblue]{hyperref}

\usepackage{cite}

\newcommand{\ket}[1]{\vert{#1}\rangle} 
\newcommand{\bra}[1]{\langle{#1}\vert} 
 
\newcommand{\bbraket}[2]{\langle\langle{#1}\vert{#2}\rangle\rangle} 
\newcommand{\proj}[1]{\ket{#1}\!\bra{#1}}
\newcommand{\op}[2]{\ket{#1}\!\bra{#2}}
\newcommand{\mean}[1]{\langle #1 \rangle}

\newcommand{\abs}[1]{\left|#1\right|} 
\newcommand{\pare}[1]{\left( #1 \right)}
\newcommand{\be}{\begin{equation}}
\newcommand{\ee}{\end{equation}}
\newcommand{\key}[1]{\left\{ #1 \right\}}
\newcommand{\cor}[1]{\left[ #1 \right]}

\newcommand{\la}{\langle}
\newcommand{\ra}{\rangle}
\newcommand{\bc}{\begin{center}}
\newcommand{\ec}{\end{center}}

\newcommand{\ben}{\begin{eqnarray}}
\newcommand{\een}{\end{eqnarray}}

\newcommand{\cH}{{\cal H}}

\newcommand{\cL}{{\cal L}}

\newcommand{\cB}{{\cal B}}
\newcommand{\cBH}{{\cal B}({\cal H})}

\newcommand{\cBaa}{\cB_{\alpha\alpha}}

\newcommand{\id}{\mathbbm{1}}

\newcommand{\oa}{\omega_{\alpha\beta\nu}(\lambda,\epsilon)}
\newcommand{\loa}{\tilde{\omega}_{\alpha\beta\nu}(\lambda,\epsilon)}

\newcommand{\mle}{\mu(\lambda,\epsilon)}

\newcommand{\cWl}{{\cal L}_{\lambda,\epsilon}}
\newcommand{\Pil}{\Pi_{12}^l}
\newcommand{\Pir}{\Pi_{12}^r}

\begin{document}


\title{Coupled activity-current fluctuations in open quantum systems under strong symmetries}


\author{D. Manzano$^{1,2}$, M.A. Mart\'inez-Garc\'ia$^{1}$, and P.I. Hurtado$^{1,2}$}
\address{$^{1}$ Departamento de Electromagnetismo y F\'{\i}sica de la Materia, Universidad de Granada, 18010 Granada, Spain.}
\address{$^{2}$  Institute Carlos I for Theoretical and Computational Physics, Universidad de Granada, 18010 Granada, Spain.}
\ead{\mailto{martinez@onsager.ugr.es, phurtado@onsager.ugr.es, manzano@onsager.ugr.es}}

\date{\today}



\begin{abstract}
Strong symmetries in open quantum systems lead to broken ergodicity and the emergence of multiple degenerate steady states. From a quantum jump (trajectory) perspective, the appearance of multiple steady states is related to underlying dynamical phase transitions (DPTs) at the fluctuating level, leading to a dynamical coexistence of different transport channels classified by symmetry. In this paper we investigate how strong symmetries affect both the transport properties and the activity patterns of a particular class of Markovian open quantum system, a three-qubit model under the action of a magnetic field and in contact with a thermal bath. We find a pair of twin DPTs in exciton current statistics, induced by the strong symmetry and related by time reversibility, where a zero-current exchange-antisymmetric phase coexists with a symmetric phase of negative exciton current. On the other hand, the activity statistics exhibits a single DPT where the symmetric and antisymmetric phases of different but nonzero activities dynamically coexists. Interestingly, the maximum current and maximum activity phases do not coincide for this three-qubits system. We also investigate how symmetries are reflected in the joint large deviation statistics of the activity and the current, a central issue in the characterization of the complex quantum jump dynamics. The presence of a strong symmetry under nonequilibrium conditions implies non-analyticities in the dynamical free energy in the dual activity-current plane (or equivalently in the joint activity-current large deviation function), including an activity-driven current lockdown phase for activities below some critical threshold. Remarkably, the DPT predicted around the steady state and its Gallavotti-Cohen twin dual are extended into lines of first-order DPTs in the current-activity plane, with a nontrivial structure which depends on the transport and activity properties of each of the symmetry phases. Finally, we also study the effect of a symmetry-breaking, ergodicity-restoring dephasing channel on the coupled activity-current statistics for this model. Interestingly, we observe that while this dephasing noise destroys the symetry-induced DPTs, the underlying topological symmetry leaves a dynamical fingerprint in the form of an intermittent, bursty on/off dynamics between the different symmetry sectors. 
\end{abstract}

\maketitle

\newpage
\tableofcontents
\newpage

\section{Introduction}
\label{s1}

The study of the statistical and thermodynamical properties of open quantum systems is one of the most fundamental problems nowadays in modern theoretical physics \cite{vinjanampathy16a,binder18a}. As the size of technological devices reduces, the understanding of quantum effects becomes crucial for the development of new solutions. Indeed, quantum effects can be engineered to increase the performance of microscopic thermal machines. Examples abound, e.g. quantum refrigerators \cite{linden10a,scully11a}, engines \cite{klimovsky15a,campisi15a,chen18a}, batteries \cite{alicki13a,campaioli18a,liu19a}, and switches \cite{tiecke14a,manzano16a}. These devices can be implemented using different nanotechnologies that are already available, including trapped ions \cite{barreiro11a,bermudez13a}, cold atoms \cite{ronzheimer13a,hild14a}, and molecular spins \cite{stamp09a,gaita-arino19a}. In most situations of interest, these systems are externally driven and operate under out-of-equilibrium conditions, making the study of nonequilibrium quantum thermodynamics crucial for the development of this emerging field. The natural framework to study this set of problems is the theory of open quantum systems \cite{breuer02a,gardiner00a}. Armed with this toolbox we can study the behavior of an open system in contact with one or several leads that drive it far from equilibrium. Such nonequilibrium systems typically evolve to a steady-state characterized by a finite current (of energy, excitations, etc.) and a well-defined stationary activity. In this way, the last years have witnessed the appearance of a number of interesting results concerning quantum nonequilibrium systems, ranging from detailed analyses of Fourier's law in one \cite{michel05a,manzano12a,znidaric11b,bulchandani20a} and several dimensions \cite{asadian13a,znidaric13b,manzano16b} to the study of quantum transport in photosynthetic compounds \cite{mohseni08a,chin10a,witt13a,manzano13a} as well as in condensed-matter systems \cite{walschaers13a,moix13a,bermudez13a}.

Fluctuations in small quantum systems play a key role as they crucially affect both their function and response to external driving. Moreover, large fluctuations (though rare) can result in drastic changes of behavior in the system of interest, and therefore the investigation of their statistics as well as the typical paths leading to them has been the focus of an intense research effort in recent years, both in the classical \cite{bertini01a,bertini02a,bertini05a,bertini06a,bertini15a,bodineau04a,derrida07a,hurtado09a,hurtado09b,hurtado10a,prados11a,hurtado11a,hurtado13a,hurtado14a,hurtado11b} and quantum \cite{esposito09a,flindt09a,garrahan10a,garrahan11a,ates12a,hickey12a,genway12a,flindt13a,lesanovsky13a,maisi14a,buca14a,zannetti14a,buca15a,znidaric14a,znidaric14b,znidaric14c,buca17a,carollo17a} realms. The mathematical framework to analyze the physics of fluctuations is large deviation theory \cite{touchette09a,garrahan10a}. The central objects of the theory are the large deviation functions (LDFs) of the different observables of interest, which allow the calculation of probabilities related to typical and not-so-typical fluctuations. The relevant observables are usually the currents (of energy, excitations, particles, etc.) characterizing nonequilibrium behavior, which comprise the time-antisymmetric sector, and the dynamical activity featuring the time-symmetric sector. LDFs for the current or the activity are of fundamental importance in nonequilibrium statistical physics as they play a role equivalent to the equilibrium free energy and related potentials, and govern macroscopic behavior out of equilibrium \cite{bertini01a,bertini02a,bertini05a,bertini06a,bertini15a,bodineau04a,derrida07a,hurtado09a,hurtado09b,hurtado10a,prados11a,hurtado11a,hurtado13a,hurtado14a,hurtado11b}. Interesting results along this research line include the existence of \emph{dynamical phase transitions} (DPTs) in the fluctuations of driven systems \cite{bertini05a, bodineau05a, harris05a, bertini06a, bodineau07a, lecomte07c, garrahan07a, garrahan09a, hurtado11a, ates12a, perez-espigares13a, harris13a, vaikuntanathan14a, mey14a, jack15a, baek15a, tsobgni16a, harris17a, lazarescu17a, brandner17a, karevski17a, carollo17a, baek17a, tizon-escamilla17b, shpielberg17a, baek18a, shpielberg18a, perez-espigares18b, chleboun18a, klymko18a, whitelam18a, vroylandt19a}, or the formulation of different fluctuation theorems for currents based on microscopic time reversibility \cite{gallavotti95a,kurchan98a,lebowitz99a,hurtado11b,andrieux09a,agarwal73a,chetrite12a,perez-espigares15a}.

Interestingly, for open quantum systems governed by a Lindblad-type master equation \cite{breuer02a,gardiner00a}, it has been recently shown that the existence of symmetries leads to different invariant subspaces and multiple (degenerate) steady-states \cite{buca12a,albert14a}. From a quantum jump (trajectory) perspective, the symmetry-induced emergence of multiple steady states is related to an underlying dynamical phase transition (DPT) in the current statistics \cite{manzano14a,manzano18a} that leads to a dynamical coexistence of different transport channels classified by symmetry. Such DPT manifests as a non-analyticity of the associated current large deviation function, a phenomenon that has been confirmed in different setups including quantum networks \cite{manzano14a} and optical switches \cite{manzano16a}; see also \cite{manzano18a}. Symmetries, and the associated dynamic phase transitions in current statistics, are very sensitive to external perturbations such as dephasing noise. However, if the noise is sufficiently weak, one can show that the existence of symmetries in the noise-free case can be inferred from the time-dependent current behavior of the (noisy) system of interest \cite{thingna16a}. Moreover, symmetries can be also manipulated by the presence of magnetic fields, resulting in a detailed control of nonequilibrium currents \cite{thingna20a}.

Up to now, research has been focused on understanding the effects of symmetries on the statistics of a single relevant observable, typically the current. A natural question hence concerns the effect of symmetries in the statistics of the dynamical activity, a time-symmetric observable of direct experimental relevance which may constraint the range of current fluctuations. Moreover, it is important to understand how symmetries are reflected in the \emph{joint} large deviation statistics of these two key observables, the current and the activity, which characterize respectively the time-antisymmetric and the time-symmetric sectors of the dynamics. In this paper we address this issue and investigate how symmetries affect both the transport properties and the activity patterns of a particular class of Markovian open quantum system, a three-qubit model under the action of a magnetic field and in contact with a thermal bath.  As expected, we find that activity constraints current fluctuations and viceversa, giving rise in particular to an activity-driven current lockdown phase induced by symmetry for this model. Interestingly, the DPT predicted at the steady state and its Gallavotti-Cohen dual are extended into lines of first order DPTs in the current-activity plane, with a nontrivial structure which depends on the transport and activity properties of each of the symmetry phases. The average current and activity of the different symmetry subspaces are analyzed in detail, as well as their dependence on the bath temperature and the external magnetic field. In addition, conditional averages such as the average current for a given value of the activity and the average activity conditioned to a given current are also explored. Finally, we also study the effect of a symmetry-breaking, ergodicity-restoring dephasing channel on the joint activity-current statistics for this model.

We structure the paper as follows. In Section \S\ref{s2} we introduce the model of interest in detail, as well as its open dynamics in terms of a Lindblad master equation for the density matrix. Section \S\ref{s3} is devoted to a symmetry analysis of the resulting dynamical equations and the associated degenerate steady states, while Section \S\ref{s35} includes quantum Monte Carlo simulations of of individual quantum trajectories. These allow us to better understand how the existence of a strong symmetry constraints the system evolution, leading to a remarkable dissipative freezing behavior, as well as to study the effect of a symmetry-breaking dephasing channel on the dynamics of qubits and the restoration of ergodicity. In Section \S\ref{s4} we introduce the counting statistics or large deviation approach to investigate the thermodynamics of quantum trajectories biased over both the current and the activity. Section \S\ref{s5} is then devoted to analyze the spectral consequences of a strong symmetry in the system, and how this leads to dynamical phase transitions in univariate large deviation functions, which affect both current and activity statistics. In Section \S\ref{s6} we investigate the joint activity-current statistics and the symmetry-induced lines of dynamical phase transitions appearing in the current-activity plane. Finally, Section \S\ref{s8} presents our conclusions and outlook for future investigation.

\section{Model and dynamics}
\label{s2}

\begin{figure}
\begin{center}
\includegraphics[scale=0.75]{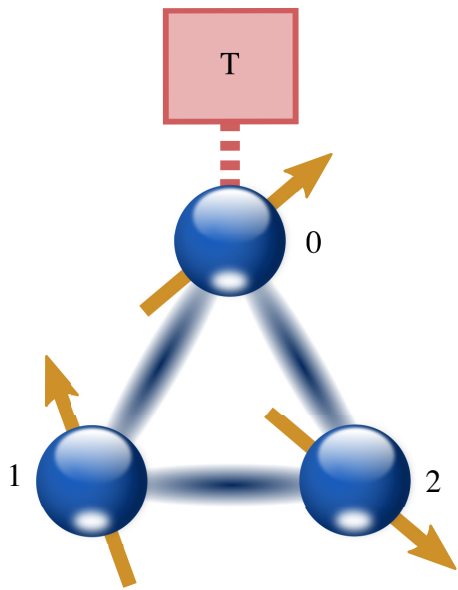}
\caption{Sketch of the three-qubit system analyzed. The blue connections represent Ising interactions while the dashed red line represents an incoherent interaction with a thermal bath.}
\end{center}
\label{fig:system}
\end{figure}

Our system consists in three spins with an $XX$ interaction term, forming an equilateral triangle as displayed in Figure \ref{fig:system}. Together with the $XX$ coupling, there is a magnetic field acting along the $Z$ direction on each spin. This kind of spin models have been broadly studied in literature, as they are relevant for a number interesting problems ranging from quantum heat conduction and Fourier's law \cite{prosen11a,manzano12a,manzano16b} to noise-assisted transport \cite{chin10a,zerah-harush18a,manzano13a} or phase transitions \cite{ajisaka14a,manzano14a,pigeon15a}, to mention just a few. Interestingly, spin systems like this one can be experimentally realized using different current technologies, including spins in semi-conductors \cite{hanson08a} and ion traps \cite{porras04a}.

The total dimension of the system is that of three-qubits. We name the pure states Hilbert space as $\cH$, having a dimension $d=2^3$. Mixed states are defined by density matrices $\rho$ that are positive trace-one operators in the space of the bounded operators $\cB(\cH)$. The Hamiltonian controlling the system coherent dynamics is 
\begin{equation}
H= \sum_{i,j =0}^2 \sigma_i^x \sigma_j^x + B_z \sum_{i=0}^2 \sigma_i^z,  
\end{equation}
where $\sigma^x_i, \sigma^z_i$ are Pauli matrices, and $B_z$ is the strength of the external magnetic field along the $Z$ direction. In addition, the system is driven by the action of a bosonic thermal bath that interacts locally with spin $0$. This bath can trigger incoherent jumps in qubit $0$ and it is modeled by the Lindblad jump operators \cite{breuer02a,gardiner00a}
\ben
L^+_0 &=& \sqrt{\Gamma n} \;\sigma_0^+ \nonumber\\
L^-_0 &=& \sqrt{\Gamma (n+1)} \; \sigma_0^-, 
\een
were $\Gamma$ is the coupling strength to the bath, $\sigma_0^{\pm} = \sigma_0^x \pm \sigma_0^y$ are raising/lowering operators acting on spin $0$, and $n=1/[\text{e}^{(\hbar B_z) /(k_B T)}-1]$ is the average number of excitations in the bath at the resonance frequency, given by a Bose-Einstein distribution at temperature $T$. For simplicity, we use units such that $\hbar=k_B=1$ throughout the paper. 

In addition to the previous ingredients, we also consider a dephasing channel acting locally and independently on all three spins. This allows us to analyze the quantum-to-classical transition at the level of trajectories, as well as the role of dephasing on symmetry-breaking and the restoration of ergodicity (see below). This dephasing channel is modeled by the jump operators  
\be
L^{\text{D}}_i= \sqrt{\gamma} \sigma_i^+ \sigma_i^-, \quad i=0,1,2
\label{eq:deph}
\ee
with $\gamma$ being the dephasing strength. The main effect of this channel is to reduce the coherences between different spins without affecting the populations in the site basis. Dephasing in nonequilibrium quantum system is known to have important effects, such as e.g. noise-enhanced transport \cite{cao09a,chin10a,manzano13a,zerah-harush18a}, current suppression \cite{scholak11a} and emergence of diffusive heat conduction \cite{manzano12a,asadian13a}. In the specific topic of symmetries and invariant subspaces, this kind of channel is often responsible of a noise-induced symmetry breaking that can collapse the multiple invariant subspaces of a Liouvillian into a single, unique steady state \cite{buca12a,manzano14a,manzano18a,thingna16a}. The global dynamics of the system is thus given by a Lindblad (or Lindblad-Gorini-Kossakowski-Sudarshan) master equation \cite{breuer02a,gardiner00a} of the form 
\be
\dot{\rho} = -i \cor{H,\rho} +  \sum_{k=\pm} L^k_0 \rho L^{k\dagger}_0 -\frac{1}{2} \key{\rho,L^{k \dagger}_0 L^k_0} 
+  \sum_{i=0}^2  L^{\text{D}}_i \rho L^{\text{D}\dagger}_i -\frac{1}{2} \key{\rho,L_i^{\text{D}\dagger} L_i^{\text{D}}} \equiv \mathcal{L} \rho  
\label{eq:me}
\ee
with $[A,B] = A B - B A$ the commutator of two operators $A$ and $B$, $\{A,B\}=AB +BA$ the anti-commutator, and $\cL$ the Liouvillian superoperator. This superoperator can be expressed in the Fock-Liouville space as a $d^2\times d^2$ complex matrix. In this way, if we have an initial state described by the density matrix $\rho(0)$, its time evolution is formally given by $\rho(t)=\exp(\cL t)\rho(0)$. According to Evan's Theorem \cite{evans79a}, a bounded system like the one discussed here should have at least one steady state, meaning that the superoperator $\cL$ defined in Eq. (\ref{eq:me}) should have at least one eigenvalue with zero real part \cite{buca12a,albert16a}. The eigenoperators associated with these null eigenvalues then correspond to the steady-state solutions of the master equation.

Master equations with local couplings have been extensively used in several fields including quantum transport \cite{manzano12a, asadian13a} and quantum thermodynamics \cite{chiara18, hewgill:prr21}. Note however that this family of equations does not arise naturally from a microscopic derivation for quantum systems locally coupled to a bath, but they can always be engineered by careful dissipation control. Indeed, in the case of a master equation microscopically derived by tracing over a bath locally coupled to a three-spin system, one would expect to obtain also global jump operators considering the collective modes of the system \cite{rivas:njp10, manzano20}. This interesting case can also exhibit symmetries leading to a  phenomenology similar to the one presented in this paper, but its analysis goes beyond the scope of this work.

\section{Symmetry analysis}
\label{s3}

In the absence of dephasing channel (i.e. $\gamma=0$), our system presents an obvious topological symmetry given by the exchange of spins $1$ and $2$, see Fig.~\ref{fig:system} and Eq. \eqref{eq:me}. Using the language of Ref. \cite{buca12a}, when $\gamma=0$ we have a \emph{strong symmetry} in the system. This is given by a unitary (and Hermitian) operator, $\pi_{12}= \frac{1}{2} \pare{ \sigma_1^x \sigma_2^x + \sigma_1^y \sigma_2^y+ \sigma_1^z \sigma_2^z  +  \id}$, being $\id$ the identity operator in the three-spins Hilbert space. This operator commutes with all the generators of the system dynamics when there is no dephasing, i.e. 
\be
\cor{\pi_{12},H}=\cor{\pi_{12},L_0^\pm}=0\, ,
\label{eq:comm}
\ee 
thus defining a strong symmetry of the dynamics \cite{buca12a}.  As the operator $\pi_{12}$ has two different eigenvalues $\key{-1,+1}$, we can spectrally-decompose the system's Hilbert space in symmetric and antisymmetric subspaces with respect to this symmetry operator, $\cH= \cH^A \oplus \cH^S$. The symmetric subspace  $\cH^S$ has dimension $d_S=6$ and it can be  spanned by the basis  $\key{S_i,i\in[1,d_S]}\equiv \key{\ket{0}_0,\ket{1}_0}\otimes\key{\ket{00}_{12},\ket{+}_{12}\equiv\frac{1}{\sqrt{2}} \pare{ \ket{01}_{12}+\ket{10}_{12}},\ket{11}_{12}}$. As expected, the basis of the subsystem formed by spins 1 and 2 in this $\cH^S$ subspace is given by the triplet states due to its symmetric nature. On the other hand, the antisymmetric subspace $\cH^A$ has dimension $d_A=2$, and we can define its basis as $\key{A_i,i\in[1,d_A]}\equiv \key{\ket{0}_0,\ket{1}_0} \otimes \key{\ket{-}_{12} \equiv\frac{1}{\sqrt{2}}\pare{ \ket{01}_{12} - \ket{10}_{12}} }$, i.e. in terms of the singlet state for the spin 1 and 2 subsystem. Note that the exchange property of the symmetry operator can be made explicit in the computational basis, i.e. $\pi_{12} = \id_0 \otimes \pare{\ket{10}\bra{01}_{12} + \ket{01}\bra{10}_{12} + \ket{11}\bra{11}_{12} + \ket{00}\bra{00}_{12} }$, with $\id_0$ the identity operator acting on spin 0. Furthermore, the Hilbert space $\cB(\cH)$ of bounded operators acting on $\cH$ can be also decomposed using the symmetry $\pi_{12}$ in the form $\cB(\cH)=\cB_{AA} \otimes \cB_{AS} \otimes \cB_{SA} \otimes\, \cB_{SS}$, with $\cB_{\alpha\beta}=\text{span} \key{\op{\alpha_i}{\beta_j}: i \in [1,d_{\alpha}], j \in [1,d_{\beta}] }$, $\alpha,\beta\in \key{S,A}$, and $S_i$, $A_i$ represent the elements of the basis of $\cH^S$ and $\cH^A$ respectively.  Note that $\cBH$ is equipped with the Hilbert-Schmidt inner product \cite{breuer02a},
\be
\bbraket{\sigma}{\rho} = \Tr(\sigma^\dagger \rho) \, , \qquad \forall \sigma,\rho \in \cBH \, ,
\label{eq:inner}
\ee
where $\Tr(\omega)$ is the trace of the operator $\omega\in\cBH$.

Due to the symmetry, the  subspaces $\cB_{\alpha\beta}$ remain invariant under the action of the Liouvillian, meaning that $\cL B_{\alpha\beta}\subset B_{\alpha\beta}$ \cite{buca12a,manzano14a,albert14a,manzano18a}, so $\cL$ can be block-decomposed into $2^2=4$ invariant subspaces. This can be easily proved by defining the left and right superoperators $\Pi_{12}^{l,r}$ such that 
\be
\Pi_{12}^{l} \rho = \pi_{12}\rho \, , \qquad \Pi_{12}^{r} \rho = \rho\pi_{12}^\dagger \, ,
\label{eq:PI}
\ee
and noticing that (a) the subspaces $\cB_{\alpha\beta}$ are the joint eigenspaces of both $\Pi_{12}^{l}$ and $\Pi_{12}^{r}$, and (b) $\cor{\Pi_{12}^{r},\cL}=0=\cor{\Pi_{12}^{l},\cL}$ due to the commutation relations \eqref{eq:comm}. In this way we obtain that, if $\rho_{\alpha\beta}\in\cB_{\alpha\beta}$, then $\cL \rho_{\alpha\beta}$ is still an eigenoperator of both $\Pi_{12}^{l,r}$ with the same eigenvalues, so that $\cL \rho_{\alpha\beta}\in B_{\alpha\beta}$. As our system is bounded, we can use now Evan's Theorem \cite{evans79a} to prove the existence of at least two fixed points of the dynamics, corresponding to the two null eigenoperators of $\cL$ in the diagonal subspaces $\cB_{\alpha\alpha}$, with $\alpha=A$ or $S$, which are the only ones to contain unit trace (physical) density matrices. Note that the subspaces $\cB_{AS}$ and $\cB_{SA}$ have no physical fixed points as they contain only zero-trace density matrices \cite{thingna20a}. In this way, there are two orthogonal steady states; if we initialize the system with a normalized (unit trace) density matrix $\rho_{\alpha}(0)\in \cB_{\alpha\alpha}$, with $\alpha=A$ or $S$, it will evolve in the long-time limit to a steady state
\be
\rho_{\alpha}^{SS} = \lim_{t\to \infty} \text{e}^{\cL t} \rho_{\alpha} (0) \in \cB_{\alpha\alpha}\, , \qquad (\alpha=A,S).
\label{eq:SS}
\ee
The existence of two steady states with typically different transport properties can be understood at the quantum trajectory level \cite{garrahan10a} as a consequence of an underlying dynamical phase transition of first-order type in the current statistics \cite{manzano14a,manzano18a}, that leads to a dynamical coexistence of different transport channels classified by symmetry (as observed above). Such dynamical coexistence has been reported in a variety of systems \cite{manzano14a,manzano18a}, including a three-spin model similar to the one described here \cite{pigeon15a}. Note also that the steady state degeneracy of open quantum systems with non-abelian symmetries has been recently addressed \cite{zhang20a}.

When restricted to the antisymmetric subspace, spins 1 and 2 stay \emph{frozen} into the singlet state, i.e. they fall into a dark (decoherence-free) state and the system dynamics is exclusively due to spin 0, connected to the bath. This is equivalent to effectively removing spins $1$ and $2$ from the total system. In this way, when restricted to the antisymmetric subspace, the system Hamiltonian can be simply written as
\be
H_A= - (\id_0 \otimes \op{-}{-}_{12})  + B_z \sigma_0^z .
\ee
For the sake of clarity, in what follows we will not make explicit identity operators (like $\id_0$ above) acting on subspaces when they can be inferred from the context. In this antisymmetric case, the steady-state density matrix is separable in the partition between qubit $0$ and qubits $1,2$. It takes the form $\rho_A^{SS}= \rho^A_{0}\otimes \proj{-}_{12}$, with
\be
\rho_0^A=
\frac{1}{1+2n}
\left(
\begin{array}{ccc}
n & & 0\\
& \\
 0
 & & 1+n\\
\end{array} 
\right),
\ee
being the thermal density matrix for one qubit in contact with a bosonic thermal bath with mean number of excitations $n$. On the other hand, the symmetric subspace interaction is described by the Hamiltonian 
\ben
\hspace{-1cm} H_S&=& \sqrt{2}\, \sigma_0^x \pare{ \op{00}{+}_{12} +\op{+}{00}_{12} + \op{+}{11}_{12} + \op{11}{+}_{12} } \\
\hspace{-1cm} &+& \left( \op{+}{+}_{12} + \op{00}{11}_{12}  +\op{11}{00}_{12} \right) 
- B_z \pare{\sigma_0^z +\op{00}{00}_{12} - \op{11}{11}_{12} }. \nonumber
\een

\section{Quantum jump trajectories}
\label{s35}

Before studying the large deviation statistics of the activity and the current in our three-qubits system (next section), we focus momentarily our interest in understanding how symmetry affects individual quantum jump trajectories. In the presence of a dephasing channel, i.e. when the dephasing rate $\gamma\ne 0$, see Eq. \eqref{eq:deph}, the exchange symmetry of the three-qubits system is broken as $\cor{L_{1,2}^D,\pi_{12}}\ne 0$. This symmetry-breaking channel restores ergodicity and leads to a unique steady-state independently of the initial state. However, for small dephasing rate the effects of the symmetry subspaces may still be observable in the transient (relaxation) behavior of the system \cite{thingna16a}.  The purpose of this section is thus to analyze the role of symmetry and its breaking in the stochastic quantum jump dynamics.

\begin{figure}
\includegraphics[width=15.5cm]{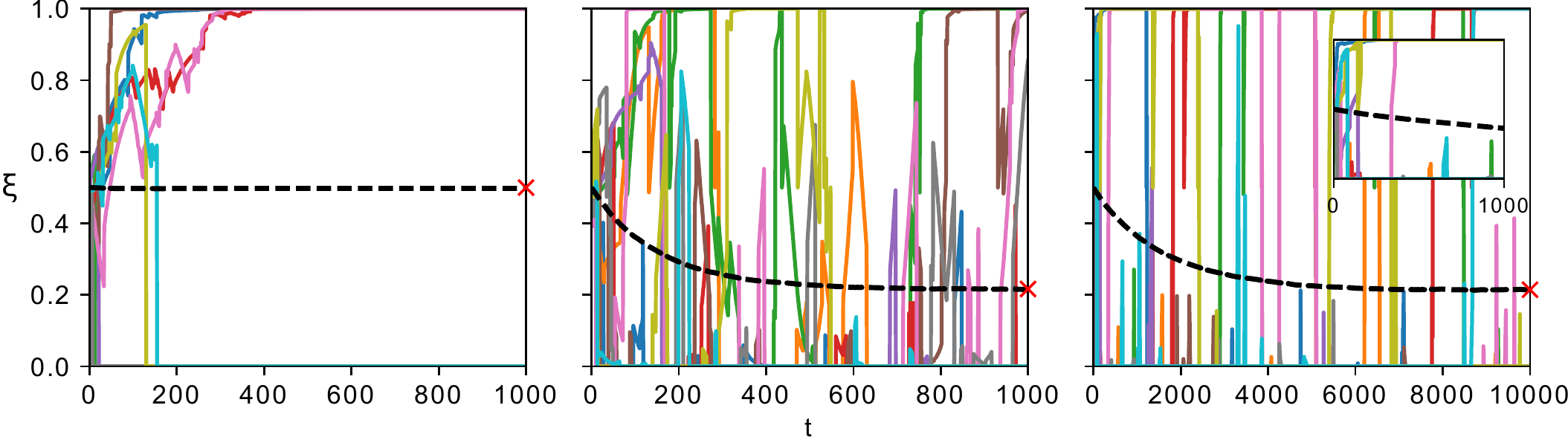}
\caption{Time evolution of the symmetry parameter $\xi(t)$ for individual quantum jump trajectories (color solid lines) and its ensemble average $\la\xi(t)\ra$  (dashed black line) obtained from quantum Monte Carlo simulations of Lindblad equation \eqref{eq:me} for the three-qubits model system. The red crosses at final times indicate the value of $\la\xi\ra$ in the steady state. The parameters are $B_z=0.5, \; \Gamma=0.1, \; n=0.5,$ and $\gamma= 0$ (left, no dephasing), $\gamma=0.01$ (center, mild dephasing), and $\gamma=0.001$ (right, weak dephasing). The inset of the right panel shows the long-time behavior for $\gamma=0.001$ (note the longer timescale as compared to the other panels). Quantum Monte Carlo ensemble averages are calculated over $10^5$ trajectories.}
\label{Fig1}
\end{figure}

To study the dynamical interplay between the different symmetry sectors as a function of dephasing, we define now a \emph{symmetry parameter} $\xi(t)$ at time $t$ in terms of the projector to the antisymmetric subspace $P_-=  \id_0 \otimes \proj{-}$. For a given pure state $\ket{\psi(t)}$ we thus define $\xi(t)\equiv \abs{P_- \ket{\psi(t)} }^2$, while for mixed states we can define its ensemble average $\la \xi(t)\ra=\Tr(P_- \rho(t))$, with $\rho(t)$ the density matrix at time $t$. In this way the symmetry parameter $\xi(t)$ captures how individual quantum jump trajectories are projected onto one of the symmetry sectors (in this case the antisymmetric one) as a function of time, quantifying the amount of \emph{symmetry selection} in the dynamical evolution. We hence generate quantum jump trajectories using a quantum Monte Carlo simulation \cite{plenio98a} of the Lindblad equation \eqref{eq:me} for our three-qubits model. Figure \ref{Fig1} shows the symmetry parameter $\xi(t)$ as measured for different quantum trajectories generated in this way, as well as its ensemble average $\la\xi(t)\ra$, as a function of time and for different values of the dephasing rate. The initial state for all quantum trajectories is taken as 
\be
\ket{\psi_0}= \frac{1}{2} \pare{\ket{0}_0+\ket{1}_0} \otimes \pare{\ket{00}_{12} + \ket{-}_{12}},
\ee
that corresponds to $\xi(0)=1/2$, i.e. a fair quantum superposition of both the symmetric and antisymmetric sectors. 

As expected, in the no-dephasing limit $\gamma=0$ (left panel in Fig.~\ref{Fig1}) both subspaces remain unmixed at the ensemble level so that the value of $\la\xi(t)\ra$ remains constant and equal to $\xi(0)$. Interestingly, however, each individual stochastic trajectory selects randomly one of the symmetry sectors, collapsing in a finite time to the corresponding subspace (making $\xi$ either $0$ or $1$) and remaining there from that time on. This remarkable behavior, known as dissipative freezing, is a particular instance of a general observation put forward in \cite{sanchez-munoz19a}, and implies a breakdown at the individual trajectory level of a conservation law associated to the symmetry operator at the ensemble level \cite{sanchez-munoz19a}. Individual quantum trajectories have equal chances to decay into either symmetry subspaces, restoring the unmixing of the symmetry sectors at the ensemble level and preserving the value of $\la\xi(t)\ra = \xi(0)$.

When dephasing noise is switched on ($\gamma>0$) the dynamics is more complex as there appears mixing between the symmetry subspaces. On one hand we find that, at the ensemble level, the average symmetry parameter $\la\xi(t)\ra$ decreases in time to reach a non-trivial steady-state value below $\xi(0)=1/2$, see center and right panels in Fig.~\ref{Fig1}, meaning that the noisy dynamics favors trajectories to collapse into the symmetric subspace. This happens because this subspace is of higher dimension ($d_S=6$ vs $d_A=2$), and hence it is entropically favored in the evolution. Interestingly, the value of $\gamma$ does not affect appreciably the steady-state value of $\la\xi\ra$, see the red cross in the center and right panels of Fig.~\ref{Fig1}. In comparison, the time required for $\la\xi(t)\ra$ to relax to its steady-state value increases as the dephasing rate $\gamma$ decreases (essentially as $1/\gamma$). The story at the level of individual quantum jump trajectories is rather different. Remarkably, for weak dephasing ($\gamma\ll 1$) the system dynamics is characterized by an intermittent, punctuated evolution, see right panel in Fig.~\ref{Fig1}, with long periods of time where the state is trapped in one of the symmetry sectors followed by quick jumps between different sectors. Such intermittent behavior is a dynamical signature of the underlying exchange symmetry \cite{thingna16a}, and remains observable as far as dephasing noise is weak. As the dephasing strength increases, this intermittent behavior tends to disappear in favor of a rapid succession of jumps between the symmetry sectors, see central panel in Fig.~\ref{Fig1}, although the overall picture is similar if time is properly rescaled by the associated relaxation timescale.


\section{Counting statistics}
\label{s4}

In Section \S\ref{s3} we have seen how the existence of a symmetry leads to multiple invariant subspaces and degenerate steady states in in our open quantum system governed by a Lindblad master equation \eqref{eq:me}. The purpose of this Section is to set the stage for trajectory statistics in open quantum systems in order to understand in subsequent sections how such symmetry affects the \emph{joint} statistical properties of two key dynamical observables, the current of excitations and the activity, for the particular case of the three-qubits model of interest in this paper. In order to do so, we use tools from large deviation theory and full-counting statistics \cite{touchette09a,garrahan10a,hurtado14a,manzano14a,manzano18a} to study the thermodynamics of quantum jump trajectories conditioned to a given total current and total total activity. The current is the key measure of transport out of equilibrium and a main token of the time-antisymmetric sector of dynamics, while the activity is a direct measure of the open quantum dynamics in the time-symmetric sector, readily accesible in experiments or simulations. Understanding their joints large-deviations statistics thus opens the door to a full characterization of the quantum jump dynamics \cite{touchette09a,garrahan10a}.

For a given quantum jump trajectory (see \cite{manzano18a} for a precise definition), the total current $Q$ after a time $t$ corresponds to the net exchange of excitons with the thermal bath. It is defined as 
\be
Q\equiv K^+ - K^-,
\ee
where $K^+$ ($K^-$) is the total number of quanta absorbed by (emitted from) the system from (to) the bath in a time interval $t$. On the other hand, the activity is just the total number of quantum jumps during such time interval,
\be
A\equiv K^+ + K^-.
\ee
Clearly, these two magnitudes are independent but correlated. In particular, the value of the activity restricts the possible values of the current since $-A\le Q \le A$, with $A\ge 0$. In addition, constraining the current to take a certain value $Q$ is expected to affect the probability distribution of the activity $A$, and viceversa. Such interplay between current and activity fluctuations is captured by their joint probability distribution \cite{touchette09a,garrahan10a,manzano18a}. To describe this joint statistics, we first introduce the reduced density matrix $\rho_{Q,A} (t)$ which is the projection of the full density matrix to the subspace defined by particular, fixed values for the total current $Q$ and activity $A$. This reduced density matrix is the solution of a current- and activity-resolved quantum master equation which can be derived from the unraveling of the Liouvillian superoperator $\cL$ in Eq. \eqref{eq:me} \cite{Derezinski08a,manzano18a}. The joint probability of observing certain values for $Q$ and $A$ after a time $t$ is thus given by $P_t(Q,A)=\Tr\cor{\rho_{Q,A} (t)}$, and scales in a large-deviation form 
\be
P_t(Q,A)\asymp \exp[+t \,G(q,a)]
\label{eq:ldp}
\ee
in the long-time limit, with $q=Q/t$ and $a=A/t$ the time-averaged (intensive) associated quantities. The symbol "$\asymp$" means asymptotic logarithmic equality, i.e. $\lim_{t\to\infty}\frac{1}{t}\ln P_t(Q,A) = G(q,a)$. The function $G(q,a)\le 0$ is the joint large deviation function (LDF) for the current and the activity, and contains all the information of the coupled fluctuations of these two central observables. 

As usual in statistical physics, working with global constraints (in this case on $Q$ and $A$) makes the problem cumbersome from a mathematical point of view (think for instance on the microcanonical ensemble in equilibrium) \cite{balescu75a,pathria09a,hurtado14a}. In order to understand the joint fluctuations it is therefore appropriate to perform a change of ensemble by introducing the Laplace transform of the reduced density matrix, 
\be
\rho_{\lambda,\epsilon}(t) =\sum_Q \sum_A \rho_{Q,A} (t) \text{e}^{-\lambda Q-\epsilon A} \, ,
\label{eq:rhole}
\ee
with $\lambda$ and $\epsilon$ different \emph{counting fields} conjugated to the current and activity, respectively, and controling their averages. By applying this transformation to the current- and activity-resolved master equation obtained from the unraveling of Eq. (\ref{eq:me}), we obtain a closed master equation for $\rho_{\lambda,\epsilon}$ \cite{manzano18a}, 
\ben
\dot{\rho}_{\lambda,\epsilon}(t) =  &-&i \cor{H,\rho_{\lambda,\epsilon}} +  \sum_{i=1}^3  L^{\text{D}}_i \rho_{\lambda,\epsilon} L^{\text{D}\dagger}_i -\frac{1}{2} \key{\rho_{\lambda,\epsilon},L_i^{\text{D}\dagger} L_i^{\text{D}}} \nonumber\\
&+& e^{-\lambda-\epsilon} L^+_0 \rho_{\lambda,\epsilon} L^{+\dagger}_0 -\frac{1}{2} \key{\rho_{\lambda,\epsilon},L^{+ \dagger}_0 L^+_0}
\label{eq:metilt} \\
&+&\, e^{+\lambda-\epsilon} L^-_0 \rho_{\lambda,\epsilon} L^{-\dagger}_0 -\frac{1}{2} \key{\rho_{\lambda,\epsilon},L^{- \dagger}_0 L^-_0}   \equiv \cL_{\lambda,\epsilon} \rho_{\lambda,\epsilon},\nonumber
\een
which defines $\cL_{\lambda,\epsilon}$, the \emph{tilted} (or \emph{deformed}) Liouvillian superoperator for the three-qubits dynamics, that no longer preserves the trace during the time evolution \cite{garrahan10a,manzano14a,manzano18a}. Interestingly, the moment generating function of the activity-current statistics is given by 
\be
Z_{\lambda,\epsilon}(t)\equiv \Tr [\rho_{\lambda,\epsilon}(t)] \, ,
\label{eq:Zle}
\ee 
which for long times also obeys a large deviation principle of the form $Z_{\lambda,\epsilon}(t)\asymp \exp[+t \, \mle]$. The function $\mle$ is nothing but the scaled cumulant generating function of the activity-current probability density function, and defines an additional LDF corresponding to the Legendre transform of $G(q,a)$, i.e 
\be
\mle=\max_{q,a}\left[G(q,a)-\lambda \, q - \epsilon \, a\right] \, ,
\label{eq:mle}
\ee
a relation equivalent to the Legendre duality between different thermodynamic potentials \cite{balescu75a,pathria09a,hurtado14a}. It can be shown \cite{manzano14a,manzano18a}, see also below, that $\mle$ is directly related to the spectral properties of the tilted superoperator $\cL_{\lambda,\epsilon}$ and the symmetry decomposition of the initial state. Note also that if we make $\lambda=0=\epsilon$ we recover the canonical (trace-preserving) Lindblad master equation (\ref{eq:me}). By inverting the Legendre transform we can conversely obtain the joint current-activity LDF $G(q,a)$ --or at least its convex envelope (see below)-- from the LDF $\mle$, i.e. $G(q,a) = \max_{\lambda,\epsilon}\left[\mle + \lambda \, q + \epsilon \, a\right]$.

On the other hand, if we make $\lambda=0$ (or $\epsilon=0$) we recover the tilted Liouvillian superoperator for the activity (or current) statistics alone. In particular, let $P_t(Q)$ be the probability of observing a total exciton current $Q$ in a time $t$. This probability obeys for long times another large deviation principle $P_t(Q)\asymp \exp[+t F(q)]$ which defines the current LDF $F(q)\le 0$ and an associated scaled cumulant generating function for the current $\theta(\lambda) = \max_q[F(q)-\lambda q] = F(q_\lambda)-\lambda q_\lambda$, with $q_\lambda$ the current associated to a given $\lambda$, solution of the equation $F'(q_\lambda)=\lambda$. Similarly, if $P_t(A)$ is the probability of observing a total dynamical activity $A$ in a time $t$, it can be shown to scale as $P_t(A)\asymp \exp[+t I(a)]$, with $I(a)\le 0$ the activity LDF such that $\zeta(\epsilon) = \max_a[I(a)-\epsilon a]=I(a_\epsilon)-\epsilon a_\epsilon$ is the scaled cumulant generating function for the activity. Here $a_\epsilon$ is the activity for a given $\epsilon$, solution of $I'(a_\epsilon)=\epsilon$. It is now easy to show that
\be
\theta(\lambda) = \mu(\lambda,0) \, , \qquad \zeta(\epsilon) =  \mu(0,\epsilon) \, .
\label{eq:LDF1}
\ee
In this way, the $k$th-order cumulants of the current and the activity are given by
\ben
\mean{q^k}_c &=& \left. - \frac{\partial^k \theta(\lambda)}{\partial \lambda^k} \right|_{\lambda\to0} = \left. - \frac{\partial^k \mu(\lambda,0)}{\partial \lambda^k} \right|_{\lambda\to0} \, , \\   
\mean{a^k}_c &=& \left. - \frac{\partial^k \zeta(\epsilon)}{\partial \epsilon^k} \right|_{\epsilon\to0} = \left. - \frac{\partial^k \mu(0,\epsilon)}{\partial \epsilon^k} \right|_{\epsilon\to0} 
\label{eq:means}
\een
which correspond to the central moments of the associated distributions up to $k=3$. Moreover, using Bayes theorem we can now define the conditional probability $P_t(Q|A)= P_t(Q,A)/P_t(A)$ to observe a total exciton current $Q$ given that the total activity is $A$, or similarly the conditional probability $P_t(A|Q)= P_t(Q,A)/P_t(Q)$ of measuring a total activity $A$ given a fixed total current $Q$. These conditional probabilities scale in the long-time limit in a large deviation form
\be
P_t(Q|A) \asymp \exp[+t G_Q(q|a)] \, , \qquad P_t(A|Q) \asymp \exp[+t G_A(a|q)] \, ,
\label{eq:ldpcond}
\ee
with
\be
G_Q(q|a) = G(q,a) - I(a) \, , \qquad G_A(a|q) = G(q,a) - F(q) 
\label{eq:ldfcond}
\ee
the associated conditional large deviation functions.

\newpage

\section{Symmetry-induced dynamical phase transitions and univariate large deviation functions}
\label{s5}

In order to analyze the role of symmetry in the joint activity-current statistics of the three-qubits system, note first that the existence of the exchange symmetry operator $\pi_{12}$ implies that the associated symmetry superoperators $\Pi_{12}^{l,r}$ defined in Eq. \eqref{eq:PI} commute with the tilted Liouville superoperator $\cL_{\lambda,\epsilon}$ of Eq. \eqref{eq:metilt}, i.e. $[\Pi_{12}^{l,r},\cL_{\lambda,\epsilon}] = 0$. Therefore there exists a complete biorthogonal basis of common left ($\loa$) and right ($\oa$) eigenoperators in $\cBH$, connecting eigenvalues of $\cWl$ to particular symmetry subspaces, such that 
\ben
\Pil\oa&=&\text{e}^{i\phi_{\alpha}}\oa \, ,  \nonumber \\
\Pir\oa&=&\text{e}^{-i\phi_{\beta}}\oa \, , \label{eq:eigenop} \\ 
\cWl\oa&=&\mu_{\nu}(\lambda,\epsilon)\oa \nonumber
\een
with $\phi_{\alpha}=0,\pi$ (and similarly for left eigenfunctions). Note that, due to the orthogonality of symmetry eigenspaces, $\Tr[\oa]\propto \delta_{\alpha\beta}$, and we introduce in what follows the normalization $\Tr[\omega_{\alpha\alpha\nu}(\lambda,\epsilon)]=1$ for simplicity. The solution to Eq. \eqref{eq:metilt} can be formally written as $\rho_{\lambda,\epsilon}(t)=\exp(+t \cWl)\rho(0)$, so a spectral decomposition of the initial density matrix in terms of the common basis yields $Z_{\lambda,\epsilon}(t)=\sum_{\alpha\nu}\text{e}^{+t\mu_\nu(\lambda,\epsilon)} \bbraket{\tilde{\omega}_{\alpha\alpha\nu}(\lambda,\epsilon)}{\rho(0)}$ for the moment generating function of the activity-current statistics. For long times we thus have
\be
Z_{\lambda,\epsilon}(t) \xrightarrow{t\to\infty} \text{e}^{+t\mu_0^{(\alpha_0)}(\lambda,\epsilon)} \bbraket{\tilde{\omega}_{\alpha_0\alpha_0 0}(\lambda,\epsilon)}{\rho(0)} \, .
\label{eq:zsum1}
\ee
Here $\mu_0^{(\alpha_0)}(\lambda,\epsilon)$ is the eigenvalue of $\cWl$ with largest real part and symmetry index $\alpha_0$ among all symmetry diagonal eigenspaces $\cBaa$ (symmetric or antisymmetric, i.e. with $\alpha=A,S$) with nonzero projection on the initial density matrix $\rho(0)$. In this way, this eigenvalue defines the Legendre transform of the joint activity-current LDF, i.e. $\mu(\lambda,\epsilon)\equiv\mu_0^{(\alpha_0)}(\lambda,\epsilon)$, see Eq. \eqref{eq:mle} above. In other words, if $\mu_0^{(\alpha)} (\lambda,\epsilon)$ is the leading eigenvalue of $\cWl$ with symmetry index $\alpha$, then
\be
\mle = \max_\alpha [\mu_0^{(\alpha)} (\lambda,\epsilon)] = \mu_0^{(\alpha_0)}(\lambda,\epsilon) \, ,
\label{eq:mumax}
\ee
with the maximum taken over all symmetry subspaces with non-zero overlap with $\rho(0)$, i.e. such that $\bbraket{\tilde{\omega}_{\alpha\alpha 0}(\lambda,\epsilon)}{\rho(0)}\ne 0$. It is interesting to note that the long time limit in Eq. \eqref{eq:zsum1} selects a particular symmetry eigenspace $\alpha_0$, 
effectively breaking at the fluctuating level the original symmetry of the three-qubits system. Remarkably, as shown in \cite{manzano14a,manzano18a}, distinct symmetry eigenspaces may dominate different fluctuation regimes, separated by first-order-type dynamical phase transitions. Moreover, the symmetry projections of the initial mixed state $\rho(0)$ can be harnessed to control both the average transport properties of the three-qubit system and its joint activity-current statistics. As an example of this mechanism applied for currents, a symmetry-controled quantum thermal switch was introduced in \cite{manzano14a} that allows the control of heat flow using initial state preparation and symmetry tools, see also \cite{manzano16a,manzano18a}.

\begin{figure}
\includegraphics[width=15.5cm]{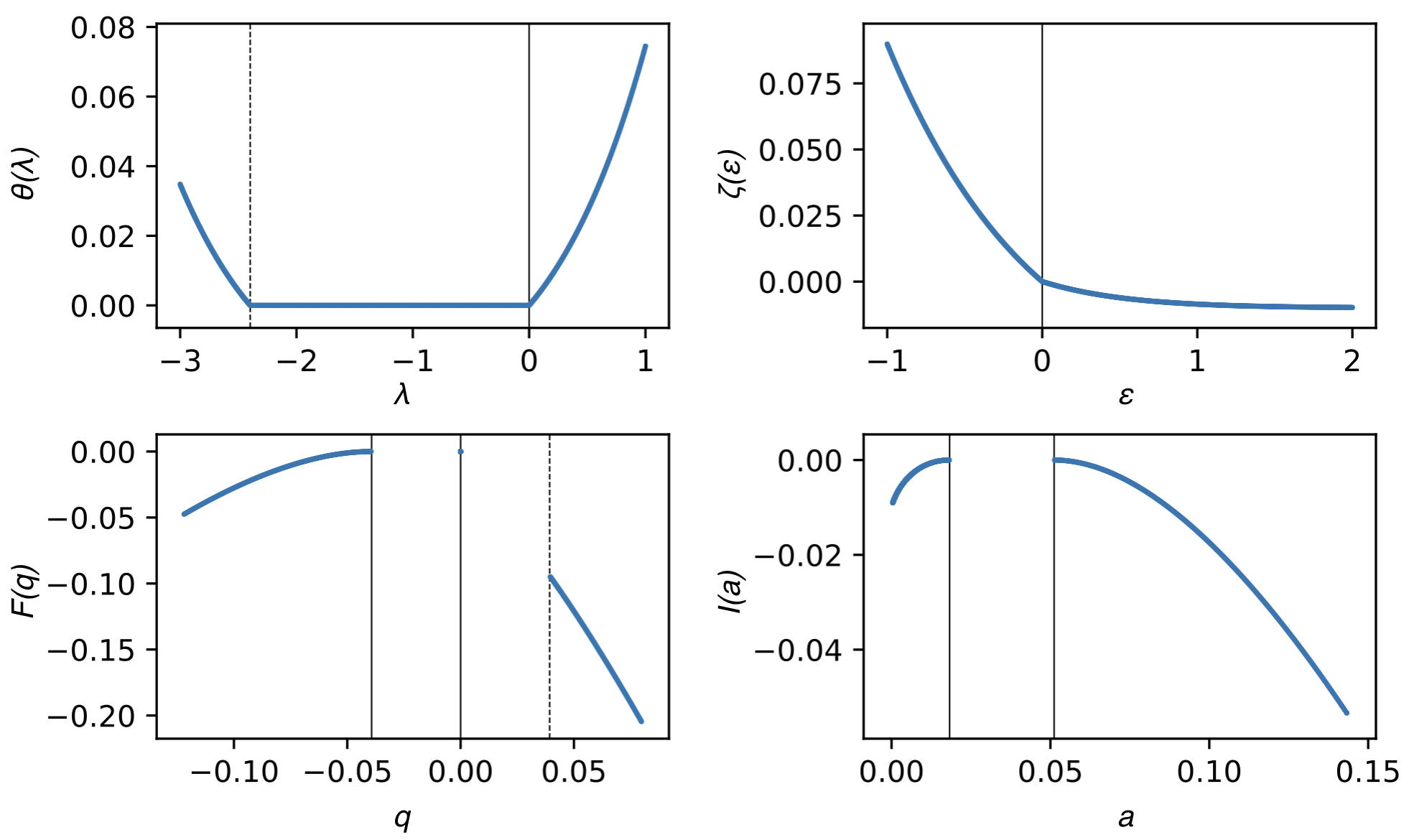}
\caption{Top row: Scaled cumulant generating functions for the exciton current, $\theta(\lambda)$ (left), and for the dynamical activity, $\zeta(\epsilon)$ (right), as a function of their respective biasing fields, for system parameters $B_z=0.5, \; \Gamma=0.1, \; \gamma= 0$, and $n=0.1$. Note the kinks in $\theta(\lambda)$ and $\zeta(\epsilon)$. Bottom row: Large deviation functions for the current, $F(q)$ (left), and for the activity, $I(a)$ (right), obtained by numerical inverse-Legendre transforming the associated scaled cumulant generating functions (top row). Note the affine or nonconvex regimes associated to the kinks above.
} 
\label{fig:mG}
\end{figure}

We next show that the existence of a symmetry such as $\pi_{12}$ implies non-analyticities in the univariate LDFs $\theta(\lambda)$ and $\zeta(\epsilon)$ associated to the current and the activity, respectively, as well as in the joint LDF $\mle$ (see next section). These non-analyticities signal dynamical phase transitions separating fluctuation regimes where the original symmetry is broken in different ways. For simplicity, we start with the current LDF $\theta(\lambda)$, see Eq. \eqref{eq:LDF1}. We hence proceed by noting that $\theta_0^{(\alpha)} (\lambda)$, the leading eigenvalue of $\cL_{\lambda,0}$ with symmetry index $\alpha$, can be expanded to first order for $\lambda \to 0$ as
\be
\theta_0^{(\alpha)} (\lambda) \approx \theta_0^{(\alpha)} (0) + \lambda \partial_\lambda \theta_0^{(\alpha)} (\lambda)|_{\lambda=0} = -\lambda \mean{q_\alpha}  \, ,
\label{eq:expa}
\ee
where we have used that $\theta_0^{(\alpha)} (0) = 0$ $\forall\alpha$ due to the existence of a well-defined steady state $\rho_\alpha^{SS}$ in each symmetry sector $\alpha$, see Eq. \eqref{eq:SS}. Moreover, $\mean{q_\alpha} = - \partial_\lambda \theta_0^{(\alpha)} (\lambda)|_{\lambda=0}$ is the average current for the steady state $\rho_\alpha^{SS}$. Using now that the cumulant generating function for the current can be written as $\theta(\lambda) = \max_\alpha [\theta_0^{(\alpha)} (\lambda)]$ \cite{manzano14a,manzano18a}, we thus find 
\be
\theta(\lambda) \underset{|\lambda|\to 0}{=} \left\{
\begin{array}{l l}
  +|\lambda| \la q_{\alpha_{\text{max}}} \ra  & \quad \text{for} \, \, \lambda \lesssim 0\\
  -|\lambda| \la q_{\alpha_{\text{min}}} \ra & \quad \text{for} \, \, \lambda \gtrsim 0
  \end{array} \right. \, ,
\label{eq:thetakink}
\ee
where $\alpha_{\text{max}}$ ($\alpha_{\text{min}}$) denotes the symmetry sector with maximal (minimal) average current $\la q_{\alpha_{\text{max}}} \ra$ ($\la q_{\alpha_{\text{min}}} \ra$) among those with nonzero overlap with $\rho(0)$. Therefore the LDF $\theta(\lambda)$ will exhibit a kink at $\lambda=0$ whenever $\la q_{\alpha_{\text{max}}}\ra \ne \la q_{\alpha_{\text{min}}} \ra$, characterized by a finite, discontinuous jump in the dynamic order parameter $q_\lambda \equiv -\theta'(\lambda)$ at $\lambda=0$ of magnitude $\Delta q_0=\la q_{\alpha_{\text{max}}} \ra - \la q_{\alpha_{\text{min}}} \ra$, a behavior reminiscent of first order phase transitions \cite{garrahan10a,manzano14a,manzano18a}. The top-left panel in Fig.~\ref{fig:mG} shows the LDF $\theta(\lambda)$ measured for our three-qubits system for a general initial state (with nonzero projections on both the symmetric and antisymmetric sectors) and a particular set of parameters ($B_z=0.5, \; \Gamma=0.1, \; \gamma= 0$, and $n=0.1$), and the presence of a kink at $\lambda=0$ is apparent, as predicted. The symmetry sector corresponding to the minimal current phase is in this case the symmetric subspace, $\la q_{\alpha_{\text{min}}} \ra = \mean{q_S}<0$, i.e. when qubits 1 and 2 are restricted to stay in any mixed state based on triplet states, see Section \S\ref{s3} above. The reason is that, interestingly, and despite the presence of a single thermal reservoir, there is a net average current of excitons \emph{from the system to the bath} in the symmetric steady state $\rho_S^{SS}$, with $\mean{q_S}=-\theta'(\lambda)|_{\lambda\to 0^+}<0$. This results from the nontrivial interplay between the Hamiltonian $XX$-interaction, the magnetic field along the $z$-direction, which induces rotation of the spins, and the thermal bath, which projects qubit 0 at a constant rate. On the other hand, as discussed in Section \S\ref{s3}, in the antisymmetric subspace spins 1 and 2 stay frozen into the singlet state, a dark (decoherence-free) state of the dynamics, effectively decoupling from the system evolution. In this symmetry sector the system thus behaves as a single qubit connected to a thermal reservoir, and the corresponding exciton current in the antisymmetric steady state $\rho_A^{SS}$ is hence $\mean{q_A}=-\theta'(\lambda)|_{\lambda\to 0^-}=0$, as deduced from the flat part of the kink at $\lambda=0^-$ of top-left panel in Fig.~\ref{fig:mG}, so that $\mean{q_A}>\mean{q_S}$.

As a result of microscopic time reversibility (since the governing Lindblad superoperator obeys a local detailed balance condition \cite{andrieux09a,agarwal73a,chetrite12a}), the probability of every quantum jump trajectory is related to the probability of its time-reversed trajectory. Both trajectories share the same value of the activity (it is a time-symmetric observable), but the current sign is reversed as it falls into the time-antisymmetric sector. Consequently, the system will obey a Gallavotti-Cohen-type fluctuation theorem for the current statistics \cite{gallavotti95a,kurchan98a,lebowitz99a} (and the joint activity-current fluctuations, see below), linking the probability of a current fluctuation with its time-reversal event. This fluctuation theorem implies that $\theta(\lambda)=\theta(\kappa-\lambda)$ for the cumulant generating function of the current, with $\kappa$ a constant related to the rate of entropy production in the system. For our three-qubits system in contact with a thermal bath characterized by an average excitation number $n$, see Section \S\ref{s2}, we have that $\kappa=\ln[n/(n+1)]$. In this way, the kink in $\theta(\lambda)$ predicted at $\lambda=0$ has a specular image at $\lambda=\kappa$, where a twin dynamical phase transition emerges in current statistics. This twin kink is confirmed in the top-right panel of Fig.~\ref{fig:mG}.

The behavior of the activity LDF $\zeta(\epsilon)$ can be analyzed in similar terms to the current. In particular, reasoning along the same lines it is easy to show that $\zeta(\epsilon)$ will show another kink around $\epsilon=0$, i.e.
\be
\zeta(\epsilon) \underset{|\epsilon|\to 0}{=} \left\{
\begin{array}{l l}
  +|\epsilon| \la a_{\beta_{\text{max}}} \ra  & \quad \text{for} \, \, \epsilon \lesssim 0\\
  -|\epsilon| \la a_{\beta_{\text{min}}} \ra & \quad \text{for} \, \, \epsilon \gtrsim 0
  \end{array} \right. \, ,
\label{eq:zetakink}
\ee
where now $\beta_{\text{max}}$ ($\beta_{\text{min}}$) denotes the symmetry sector with maximal (minimal) average activity $\la a_{\beta_{\text{max}}} \ra$ ($\la a_{\beta_{\text{min}}} \ra$) among those with nonzero overlap with $\rho(0)$. For the particular case of the three-qubits system with exchange symmetry studied in this paper, the maximum current and maximum activity subspaces do not coincide, as the maximum current subspace corresponds to the antisymmetric one while the maximum activity subspace corresponds to the symmetric sector, i.e. $\mean{a_{\beta_{\text{max}}}}=\mean{a_S}$ and $\mean{a_{\beta_{\text{min}}}}=\mean{a_A} < \mean{a_S}$. We stress however that this is not a necessary condition in general cases (for other open quantum systems with different symmetries, and/or different joint observables other than the current and the activity). Top-right panel in Fig.~\ref{fig:mG} shows the measured $\zeta(\epsilon)$ for the three-qubits system and the same particular set of parameters, confirming the presence of this additional kink. Notice also that, as opposed to the current, the dynamical activity is a time-symmetric observable which does not change sign upon time-reversal (indeed the activity is a positive-definite observable). Therefore no twin kink in $\zeta(\epsilon)$ is expected in this case, as confirmed in the top-right panel of Fig.~\ref{fig:mG}.

As an interesting corollary, note that the kinks in the current LDF $\theta(\lambda)$ can only happen \emph{out of equilibrium}, disappearing in equilibrium. In particular, the average currents for the multiple steady states are by definition zero in equilibrium, $\mean{q_\alpha} = 0$ $\forall \alpha$, so no symmetry-induced first-order dynamical phase transition appears in $\theta(\lambda)$ at $\lambda=0$ in equilibrium. On the other hand, the average activities of the different symmetry subspaces can be still different even when the system is in equilibrium, so the kink in the activity LDF $\zeta(\epsilon)$ and the associated activity dynamical phase transition may still be present in equilibrium.

\begin{figure}
\includegraphics[width=15.5cm]{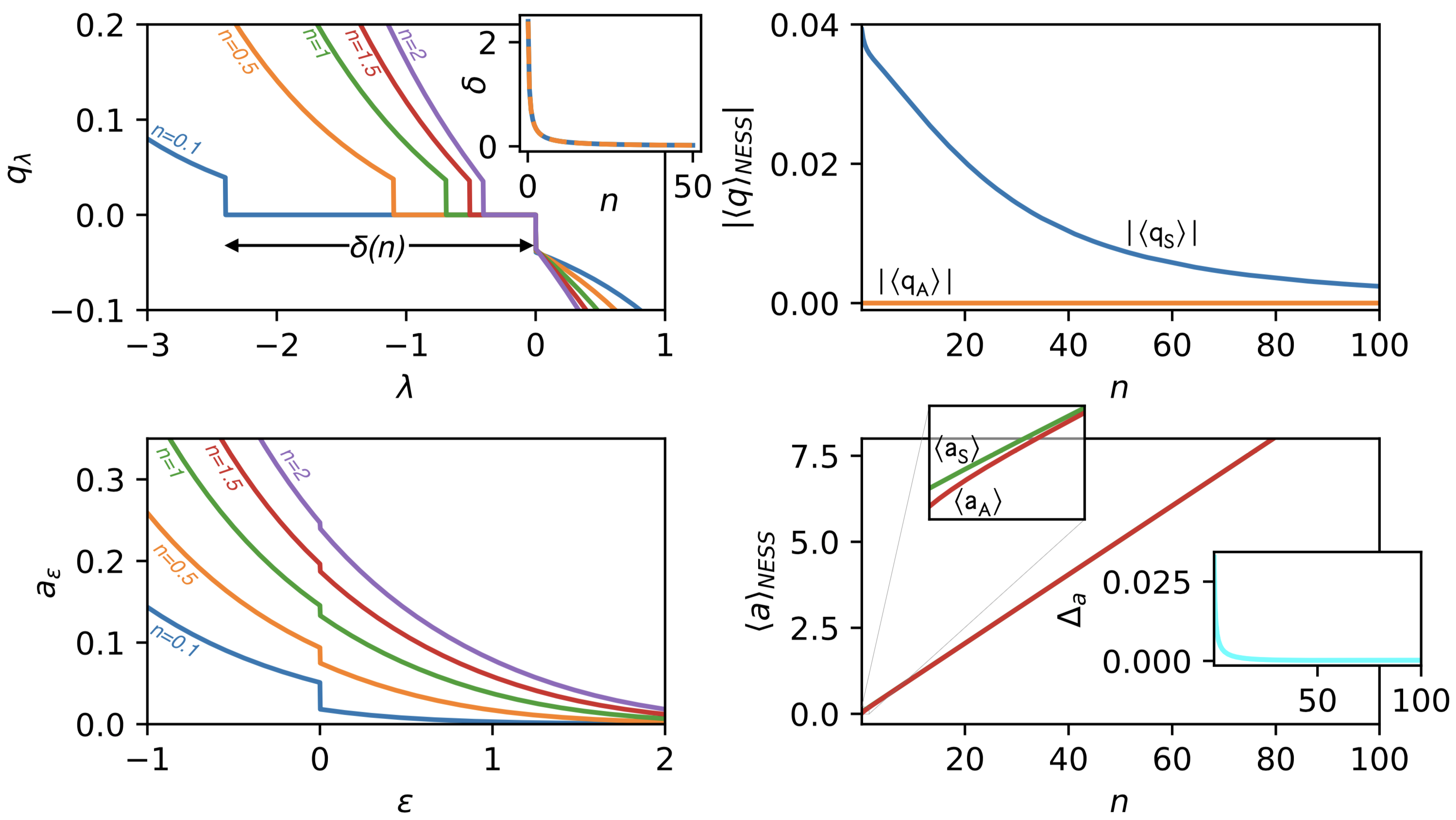}
\caption{Dependence of different observables with the bath average number of excitations $n$ for $B_z=0.1, \; \Gamma=0.1$, and $\gamma= 0$. Top left: Current $q_\lambda=-\theta'(\lambda)$ as a function of the bias parameter $\lambda$ for different values of $n$. The inset shows the width of the discontinuous gap $\delta$ as a function of $n$. The dashed line is the prediction $\delta(n)=-\kappa=\ln[(n+1)/n]$. Top right: Absolute value of the average current in the symmetric ($|\mean{q_S}|$) and antisymmetric ($|\mean{q_A}|$) subspaces as a function of $n$. Bottom left: Activity $a_\epsilon=-\zeta'(\epsilon)$ as a function of the bias parameter $\epsilon$ for different values of $n$. Bottom right: Average activity in the symmetric ($\mean{a_S}$) and antisymmetric ($\mean{a_A}$) subspaces as a function of $n$. The top inset shows a zoom to the low-$n$ behavior, where both average activities are clearly different. The bottom inset shows the difference $\Delta a = \mean{a_S} - \mean{a_A}$ vs $n$. 
} 
\label{fig:qn}
\end{figure}

We may now obtain the current LDF $F(q)$ from the inverse Legendre transform of $\theta(\lambda)$, i.e. $F(q)=\max_{\lambda}[\theta(\lambda) + q\lambda]$. It is then easy to show \cite{touchette09a} that the twin kinks in $\theta(\lambda)$ correspond to two different current intervals, $|q|\in(0,|\mean{q_S}|]$ (related by time-reversibility, $q\leftrightarrow -q$) where $F(q)$ is affine or nonconvex \cite{manzano14a}, as corresponds to a multimodal current distribution $P_t(Q)$ reflecting the dynamical coexistence of multiple transport channels (or steady states) classified by symmetry. Bottom-left panel in Fig.~\ref{fig:mG} shows $F(q)$ as obtained by numerical inverse Legendre transform of $\theta(\lambda)$ in the top-left panel of the same figure. Note that the maximum current regime reduces to a single point in $q$-space as $\mean{q_{\alpha_{\text{max}}}}= \mean{q_A}=0$. In this way, in order to sustain a given current fluctuation $q$ such that $|q|\notin(0,|\mean{q_S}|]$, the open quantum system breaks the original symmetry and selects the particular symmetry sector that maximally facilitates a given current fluctuation: the statistics during a current fluctuation with $|q|\ge |\la q_{S} \ra|$ is dominated by the symmetric subspace, whereas for zero current $q=0$ the antisymmetric subspace prevails. Moreover, for currents $|q|\in(0,|\mean{q_S}|]$ the dominant quantum jump trajectory spends some time $t_0=pt$ ($p<1$) in the symmetric sector and a complementary time $t-t_0=t(1-p)$ in the antisymmetric subspace, with $p=|q/\mean{q_S}|$, a sort of dynamical Maxwell-like construction \cite{perez-espigares18a}. Equivalent arguments hold for the activity LDF $I(a)$, shown in the bottom-right panel of Fig.~\ref{fig:mG}, which exhibits an affine or nonconvex regime for activities $a\in[\la a_{\beta_{\text{min}}} \ra,\la a_{\beta_{\text{max}}} \ra] = [\mean{a_A},\mean{a_S}]$ as a result of the kink in $\zeta(\epsilon)$ at $\epsilon=0$, with $\mean{a_A} < \mean{a_S}$. Note however that, due to the time-symmetric character of the activity, no twin dynamical phase transition is expected in this case (note also that $a\ge 0$ in all cases).

\begin{figure}
\includegraphics[width=15.5cm]{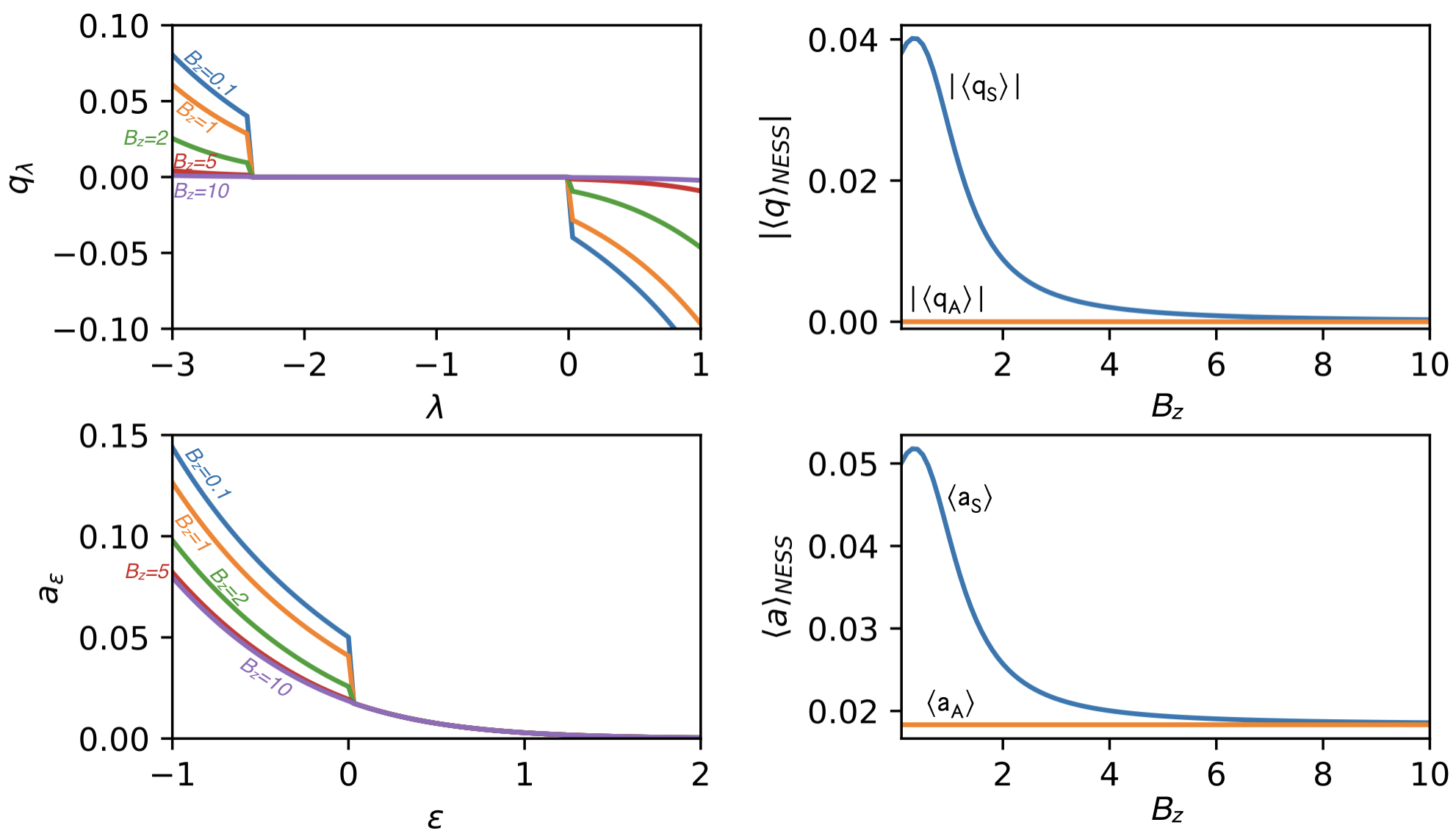}
\caption{Effect of the magnetic field intensity on different magnitudes for $n=0.1, \; \Gamma=0.1$, and $\gamma= 0$. Top left: Current $q_\lambda=-\theta'(\lambda)$ as a function of the bias parameter $\lambda$ for different magnetic fields $B_z$. Top right: Absolute value of the average current in the symmetric ($|\mean{q_S}|$) and antisymmetric ($|\mean{q_A}|$) subspaces as a function of $B_z$. Bottom left: Activity $a_\epsilon=-\zeta'(\epsilon)$ as a function of the bias parameter $\epsilon$ for different values of $B_z$. Bottom right: Average activity in the symmetric ($\mean{a_S}$) and antisymmetric ($\mean{a_A}$) subspaces as a function of $B_z$. 
} 
\label{fig:qB}
\end{figure}

Next we study how the average current and activity of the different symmetry-classified steady states behave as a function of the average number of excitations $n$ in the thermal bath, see top-right and bottom-right panels in Fig.~\ref{fig:qn}. We first note that, while the average activity in both the symmetric and antisymmetric sectors increases with $n$ (bottom-right panel in Fig.~\ref{fig:qn}), the absolute value of the average current in the symmetric sector decreases instead with $n$ (the current in the antisymmetric sector vanishes $\forall n$ as explained above). In this way the average current in the symmetric sector recedes from its activity-related bound, $-\mean{a_S}\le \mean{q_S}\le \mean{a_S}$, as $n$ increases. The activity of both the symmetric and antisymmetric steady states grows linearly with $n$ for large enough $n$, although the activity of the symmetric sector is always (slightly) above the one of the antisymmetric sector. Note however that activity differences between both symmetry sectors are only apparent for low values of the bath average excitation number $n$, see top inset in the bottom-right panel of Fig.~\ref{fig:qn}. Moreover, the differences in the transport and activity patterns between the symmetric and antisymmetric sectors of the dynamics tends to vanish as $n$ increases, thus reducing the range of controllability of the quantum current and activity, possible by tuning the symmetry projections of the initial state \cite{manzano14a,manzano16a,manzano18a}. We also studied the dependence with $n$ of the typical current $q_\lambda=-\theta'(\lambda)$ for a bias parameter $\lambda$, and the typical activity $a_\epsilon=-\zeta'(\epsilon)$ for bias $\epsilon$, see top left and bottom-left panels in Fig.~\ref{fig:qn}, respectively. The discontinuities in both $q_\lambda$ and $a_\epsilon$ around the kinks in their respective LDFs are apparent. Moreover, the width of the discontinuous gap in $q_\lambda$, denoted here as $\delta$ and associated to the regime in $\lambda$-space dominated by the antisymmetric sector, quickly decreases with $\lambda$. This gap is related to the $\lambda$-distance between a given event and its time-reversal, and is simply given by $\delta = -\kappa = \ln[(n+1)/n]$, see inset in top-left panel of Fig.~\ref{fig:qn}. In comparison, the size of the discontinuous jumps in $q_\lambda$, related to the difference in average currents between the symmetric and antisymmetric sectors, decreases at a much slower pace with $n$. For the activity jump, on the other hand, the decrease with $n$ is much faster.

Fig.~\ref{fig:qB} explores the dependence of the same observables with the strength of the external magnetic field $B_z$. Interestingly, both the average current and the average activity of the symmetric steady state depend \emph{non-monotonously} on the strength of the magnetic field, see top-right and bottom-right panels in Fig.~\ref{fig:qB}, while $B_z$ does not affect the values of $\mean{q_A}$ and $\mean{a_A}$. In particular, there exists a particular (common) value of $B_z$ where the differences $\mean{q_S}-\mean{q_A}$ and $\mean{a_S}-\mean{a_A}$ are maximal. This may be used to optimize the range of controllability of the transport and activity properties of the three-qubits systems \cite{manzano14a,manzano16a,manzano18a}. Due to the activity constraint on the current, $-\mean{a_S}\le \mean{q_S}\le \mean{a_S}$, the decrease of $\mean{a_S}$ with $B_z$ forces $\mean{q_S}$ to diminish also as the magnetic field increases, a sort of enslaved behavior. Note also that, contrary to what happens when $n$ is changed, see Fig.~\ref{fig:qn}, the width of the discontinuous gap $\delta$ in $q_\lambda$ does not depend on the magnetic field intensity $B_z$, see top-left panel in Fig.~\ref{fig:qB}, indicating that $B_z$ does not affect the rate of entropy production in the system.

\section{Joint activity-current fluctuations}
\label{s6}

\begin{figure}
\center{\includegraphics[width=9cm]{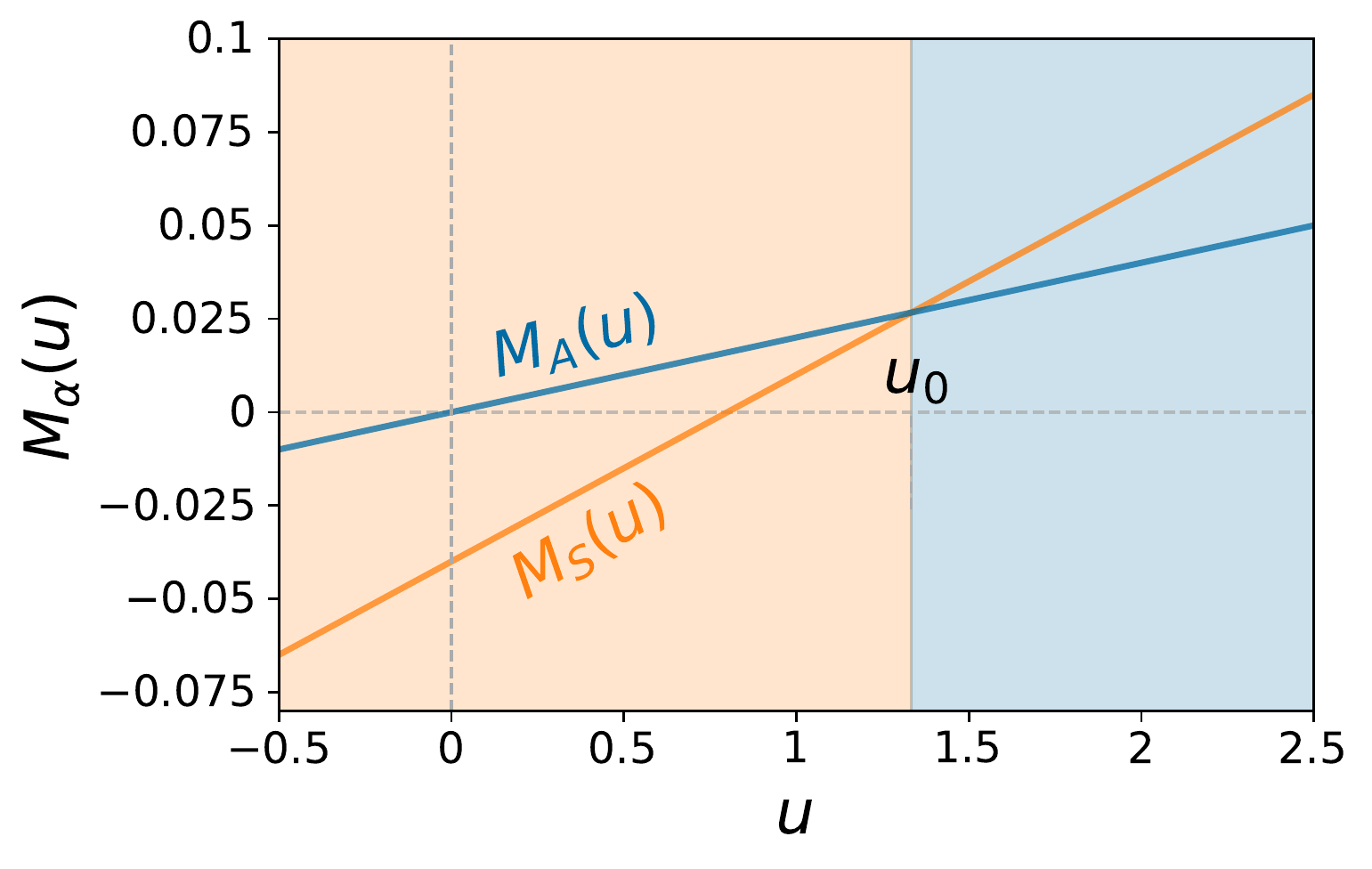}}
\caption{Linear function $M_\alpha(u)= \la q_\alpha\ra + u\la a_\alpha\ra$, with $\alpha=S,A$, as a function of $u=\epsilon/\lambda$ for the three-qubits system and a particular set of parameters. Note the existence of a crossing point $u_0$, see Eq. \eqref{eq:u0}, such that $M_A(u)<M_S(u)$ for $u>u_0$ while $M_S(u)<M_A(u)$ for $u<u_0$. This implies that for $\lambda>0$, where $\mle \underset{\lambda,\epsilon\to 0}{=}-\lambda \min_\alpha M_\alpha(u)$,  activity-current fluctuations are dominated by the antisymmetric sector for $u>u_0$ (blue shaded area), while the symmetric subspace prevails for $u<u_0$ (red shaded area). This prevalence behavior is inverted for $\lambda<0$.} 
\label{fig:Mu}
\end{figure}

We next turn to investigate the joint large deviation function $\mle$, a bivariate LDF for which the arguments are similar to those discussed in previous section but some care is needed. Proceeding as in Section \S\ref{s5}, see e.g. Eq. \eqref{eq:expa}, we first note that the leading eigenvalue of the tilted superoperator $\cWl$ with symmetry index $\alpha$, denoted as $\mu_0^{(\alpha)} (\lambda,\epsilon)$, can be expanded to first order for $\lambda,\epsilon \to 0$ as
\be
\mu_0^{(\alpha)} (\lambda,\epsilon) \underset{|\lambda|,|\epsilon|\ll 1}{\approx} \mu_0^{(\alpha)} (0,0) + \lambda \partial_\lambda \mu_0^{(\alpha)} (\lambda,0)|_{\lambda=0} + \epsilon \partial_\epsilon \mu_0^{(\alpha)} (0,\epsilon)|_{\epsilon=0} = -\lambda \mean{q_\alpha} - \epsilon \mean{a_\alpha} \, ,
\label{eq:expamu}
\ee
where we have used in the second equality that $\mu_0^{(\alpha)} (0,0) = 0$ $\forall\alpha$ due to stationary conditions in each symmetry subspace, as above. Using now that the joint LDF $\mle = \max_\alpha [\mu_0^{(\alpha)} (\lambda,\epsilon)]$, we thus find that $\mle \underset{\lambda,\epsilon\to 0}{=} \max_\alpha [-\lambda \mean{q_\alpha} - \epsilon \mean{a_\alpha}]$. Since we are working in the limit where $\lambda,\epsilon \to 0$, both at comparable rates (otherwise the faster-decaying biasing field would be effectively zero in this limit, thus reducing the discussion to that in the previous section for univariate LDFs), we can write $\epsilon = u\lambda$ with $u$ some proportionality constant so 
\be
\mle \underset{\lambda,\epsilon\to 0}{=} \max_\alpha [-\lambda M_\alpha(u)] = \left\{ 
\begin{array}{ll}
\displaystyle -\lambda \min_\alpha M_\alpha(u) & (\lambda>0) \\
\displaystyle |\lambda| \max_\alpha M_\alpha(u) & (\lambda<0)
\end{array} \right. \, ,
\label{eq:mukink}
\ee
where we have defined a linear function 
\be
M_\alpha(u)=\mean{q_\alpha} + u \mean{a_\alpha}
\label{eq:Mau}
\ee
of slope $\mean{a_\alpha}$ and intercept $\mean{q_\alpha}$, characteristic of each symmetry subspace with index $\alpha$. Fig.~\ref{fig:Mu} shows a sketch of $M_\alpha(u)$ for the three-qubits system studied here and the particular set of parameters used e.g. in Fig.~\ref{fig:mG} ($B_z=0.5, \; \Gamma=0.1, \; \gamma= 0 , \; n=0.1$), with $\alpha=S$ (symmetric subspace) and $\alpha=A$ (antisymmetric sector). In general, there exists a threshold value 
\be
u_0=-\frac{\la q_S \ra - \la q_A \ra}{\la a_S \ra - \la a_A \ra} \, ,
\label{eq:u0}
\ee
defined by the equality $M_S(u_0)=M_A(u_0)$, such that $M_A(u)<M_S(u)$ for $u>u_0$ while $M_S(u)<M_A(u)$ for $u<u_0$, see Fig.~\ref{fig:Mu}. In this way, the symmetry sector dominating the joint activity-current fluctuations near the steady state will depend on how we approach the origin in $(\lambda,\epsilon)$-space, i.e. on the particular value of parameter $u=\epsilon/\lambda$. For $\lambda>0$ we have that $\mle \underset{\lambda,\epsilon\to 0}{=}-\lambda \min_\alpha M_\alpha(u)$, so for $u>u_0$ the antisymmetric sector dominates activity-current fluctuations, while for $u<u_0$ the symmetric subspace is responsible of activity-current fluctuations, see Fig.~\ref{fig:Mu}. Conversely, for $\lambda<0$ we have that $\mle \underset{\lambda,\epsilon\to 0}{=} |\lambda| \max_\alpha M_\alpha(u)$, so for $u>u_0$ the symmetric sector prevails, while the antisymmetric subspace dominates for $u<u_0$. Therefore $\mle$ exhibits a kink line (i.e. with discontinuous derivative) near $(\lambda,\epsilon)\to 0$, with local slope $u_0$ near the origin, such that the symmetric sector dominates \emph{below} this line while the antisymmetric subspace prevails \emph{above} it.

\begin{figure}
\includegraphics[width=15.5cm]{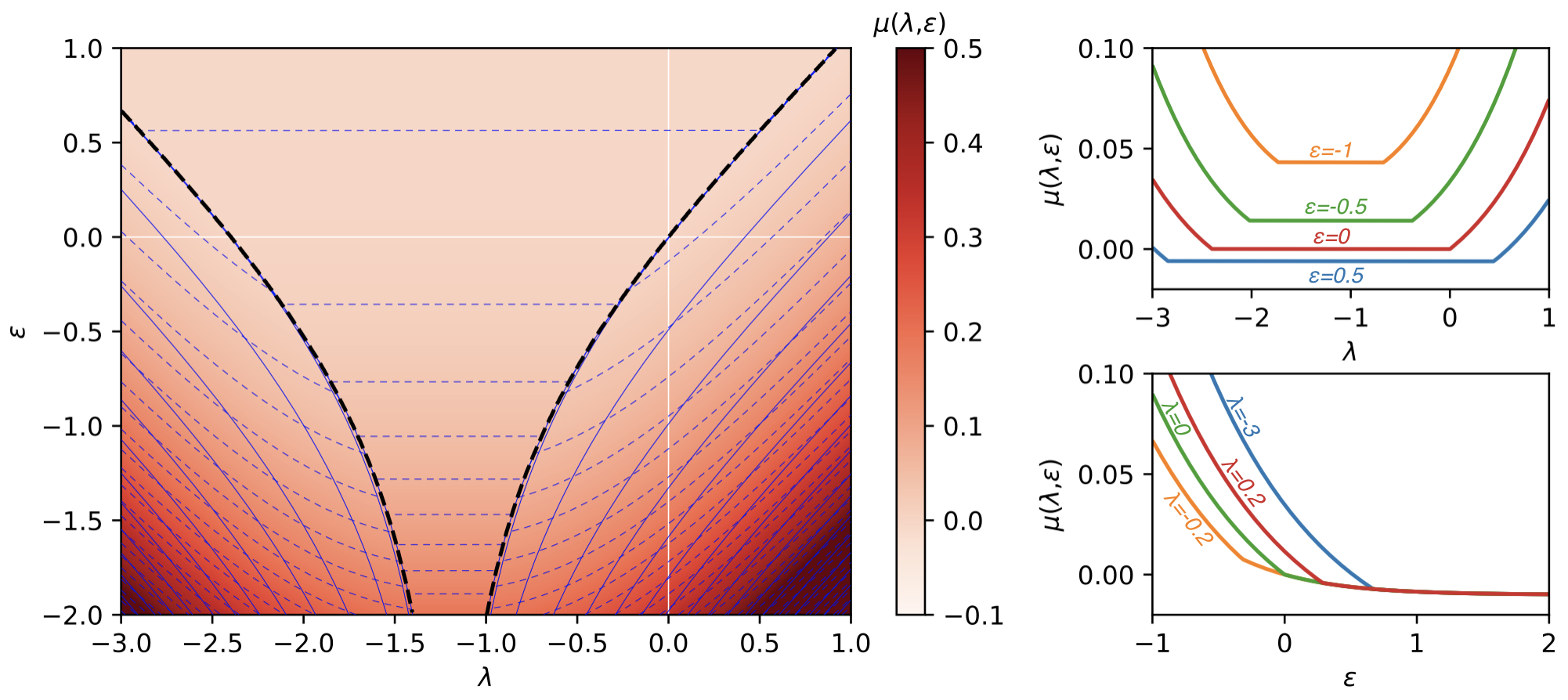}
\caption{Joint current-activity scaled cumulant generating function. Left: Color map of $\mle$ as a function of $\lambda$ and $\epsilon$. The dashed thick black lines mark the twin kinks in the LDF and signal the non-analyticity of the derivative. The white lines mark the zero values of $\lambda$ and $\epsilon$. Note the non-trivial slope $u_0$ of the kink line around $\lambda=0=\epsilon$, see Eq.~\eqref{eq:u0}. Thin solid blue lines represents isolines of constant $q \in \cor{-0.74,\;0.53}$ while thin dashed blue lines are isolines of constant $a \in \cor{0.005,\;0.59}$) Right top: $\mle$ as a function of $\lambda$ for different values of $\epsilon$. Right bottom: $\mle$ as a function of $\epsilon$ for different values of $\lambda$. In all panels the system parameters are $B_z=0.5, \; \Gamma=0.1, \; \gamma= 0$, and $n=0.1$.} 
\label{fig:mle}
\end{figure}

Fig.~\ref{fig:mle} (left panel) shows the measured LDF $\mle$ for the three-qubits model as a function of $\lambda$ and $\epsilon$, for parameters $B_z=0.5, \; \Gamma=0.1, \; \gamma= 0$ and $n=0.1$, as in previous plots. Interestingly, the infinitesimal kink segment predicted around $\lambda=0=\epsilon$ of local slope $u_0$ is confirmed, and extends over the entire $(\lambda,\epsilon)$-plane into a line of first-order dynamical phase transitions along which dynamical coexistence of the different (symmetric and antisymmetric) transport channels appears. Furthermore, due to the microscopic time-reversibility of the open quantum dynamics \cite{andrieux09a,agarwal73a,chetrite12a}, the joint activity-current LDF $G(q,a)$ obeys a Gallavotti-Cohen-type fluctuation theorem along the current (time-antisymmetric) axis \cite{gallavotti95a,kurchan98a,lebowitz99a}, which for the Legendre-dual LDF can be simply written as $\mle = \mu(\kappa-\lambda,\epsilon)$, where $\kappa=\ln[n/(n+1)]$ has been defined above. This immediately implies a twin kink branch in the $(\lambda,\epsilon)$-plane signaling a twin line of dynamical phase transitions, as confirmed in the left panel of Fig.~\ref{fig:mle}. In this way, the $(\lambda,\epsilon)$-plane is divided into two regions, an \emph{inner zone} where the associated joint activity-current fluctuations are dominated by the antisymmetric subspace, and an \emph{outer region} where the symmetric sector prevails. The presence of a double kink along the $\lambda$-axis and a single kink along the $\epsilon$-axis can be confirmed by representing constant-$\epsilon$ and constant-$\lambda$ slices of the LDF $\mle$, see respectively top-right and bottom-right panels in Fig.~\ref{fig:mle}. 

\begin{figure}
\includegraphics[width=15.5cm]{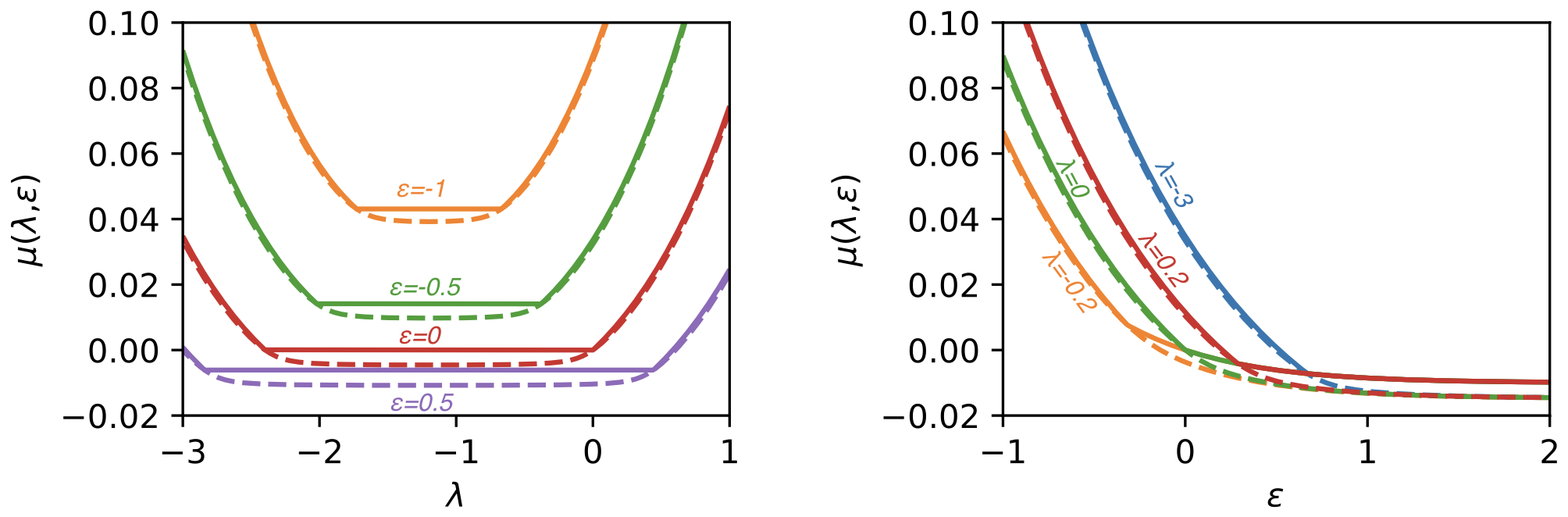}
\caption{Role of dephasing noise. Left: The large deviation function $\mle$ as a function of $\lambda$ for different values of $\epsilon$. Right: The large deviation function $\mle$ as a function of $\epsilon$ for different values of $\lambda$. In both panels the solid lines represent the no-dephasing case ($\gamma=0$) while the dashed lines represent the effect of a small dephasing  ($\gamma=0.01$). The parameters are $B_z=0.5, \; \Gamma=0.1$ and $n=0.1$.}
\label{fig:mledeph}
\end{figure}

As discussed in Section \S\ref{s35}, a noisy dephasing channel acting on all qubits ($\gamma\ne 0$, see Eq. \eqref{eq:deph} and Section \S\ref{s2}) breaks the exchange symmetry of the original system, restoring global ergodicity and leading to an unique steady state \cite{evans79a,buca12a}. The presence of this symmetry-breaking dephasing noise then immediately implies the disappearance of the symmetry-induced dynamical phase transitions and the kinks in $\mle$ described above, as well as the kinks in the univariate LDFs $\theta(\lambda)$ and $\zeta(\epsilon)$. This is illustrated in Fig.~\ref{fig:mledeph}, where constant-$\epsilon$ and constant-$\lambda$ slices of the LDF $\mle$ are represented, both in the absence ($\gamma=0$) and in the presence ($\gamma\ne 0$) of dephasing channel. The existence of the kink is apparent in the case with no dephasing (solid lines), but it disappears when $\gamma\ne 0$ (dashed lines). It is also remarkable that, in the symmetric-subspace phase, $\mle$ is very similar with and without dephasing, while in the antisymmetric-subspace regime the difference is appreciable, e.g. $\mle$ vs $\lambda$ for constant-$\epsilon$ is flat in this antisymmetric regime when $\gamma=0$, but it seems to continue analytically the form of $\mle$ in the symmetric case when $\gamma\ne 0$, see left panel in Fig.~\ref{fig:mledeph}. This happens because the action of the dephasing channel changes dramatically the symmetry properties of the system, but not its transport properties. The observed behavior is another indication that, once the symmetry is broken by the dephasing channel, the dominant subspace in the dynamical evolution is the symmetric one (due to entropic reasons), as already discussed in Section \S\ref{s35}.

\begin{figure}
\includegraphics[width=15.5cm]{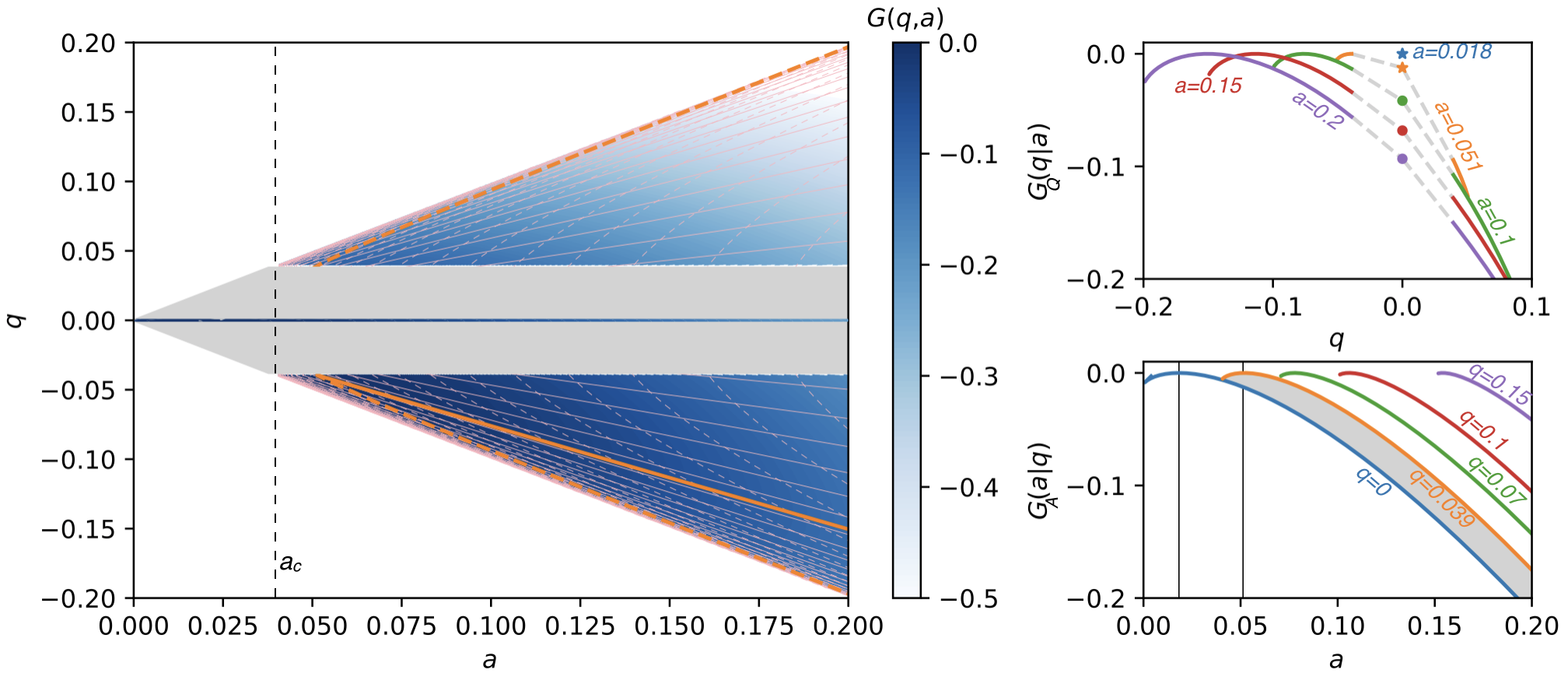}
\caption{Joint activity-current statistics. Left: Color map of $G(q,a)$ as a function of $q$ and $a$. The grey areas represent the affine or nonconvex regions that cannot be recovered by inverse Legendre-transforming $\mle$. Thin solid grey lines represents the isolines of constant $\lambda\in[-4,2]$, while thin dashed grey lines are isolines of constant $\epsilon\in[-2,4]$. The thick orange solid line represents $\mean{q_a}$ (corresponding to $\lambda=0$), while the thick orange dashed line is $\mean{a_q}$ (corresponding to to $\epsilon=0$). Right top: Conditional current LDF $G_Q(q|a)$ for different, fixed activities $a$. The grey dashed lines represents the affine or nonconvex regions that cannot be recovered from $\mle$. Right bottom: Conditional activity LDF $G_A(a|q)$ for varying, fixed currents $q$. The grey area represents the affine or nonconvex region that cannot be recovered from $\mle$.  In all panels, the parameters are $B_z=0.5, \; \Gamma=0.1, \; \gamma= 0$ and $n=0.1$.} 
\label{fig:gqa}
\end{figure}

We can now obtain the joint current-activity LDF $G(q,a)$ by inverse Legendre transforming $\mle$, see left panel in Fig.~\ref{fig:gqa}.  First, it is important to note that the activity constraint on the current, $-a\le q\le a$, has significant consequences in the joint activity-current statistics. For instance, the constraint implies that the absolute value of the current cannot be larger than the activity under any circumstances, so $G(q,a)\to -\infty$ for any $|q|>a$ and therefore $G(q,a)$ takes finite values only in a triangle defined by the constraint $|q|\le a$.  On the other hand, since both the symmetric and antisymmetric subspaces have definite values of the average current, $\mean{q_A}=0$ and $\mean{q_S}<0$ respectively, the constraint $-a\le q\le a$ implies that there exists a critical value for the activity $a_c=|\mean{q_S}|$ such that the current for any activity $a<a_c$ can only be $q=\mean{q_A}=0$. This immediately implies that $G(q=0,a<a_c)=0$ while $G(q\ne0,a<a_c)= -\infty$, see main plot in Fig.~\ref{fig:gqa}. Remarkably, this clear-cut observation opens up a new route to control quantum transport: by biasing the activity of our three-qubits system below the critical activity $a_c$, one is able to shut down completely the exciton current in the system, since in this fluctuation regime the antisymmetric sector prevails. This novel \emph{activity-driven current lockdown regime} is enabled by symmetry, and suggests unexplored quantum control strategies. In addition, proceeding as in Section \S\ref{s5}, one can easily show \cite{touchette09a,manzano14a} that the twin kink branches in $\mle$ correspond (after the inverse Legendre transform) to two different regions in the $(q,a)$-plane where the $G(q,a)$ is affine or nonconvex, signaling the dynamical coexistence of the two different transport channels in these current-activity regions. These affine or nonconvex zones are clearly visible in the main panel of Fig.~\ref{fig:gqa}, and comprise two well-defined bands $(-|\mean{q_S}|,0)\cup [a_c,+\infty)$ and $(0,|\mean{q_S}|)\cup [a_c,+\infty)$ in $(q,a)$-space. Note that between these two zones there is a $q=0$ line $\forall a$ corresponding to the antisymmetric manifold. Note also that negative currents are more probable than positive ones, see color legend in left panel of Fig.~\ref{fig:gqa}, and hence the average current is negative. For clarity, we have represented by a thick orange solid curve the isoline $\lambda=0$ in the $(q,a)$-plane, that marks the average current $\mean{q_a}$ for a given activity $a$, while the dashed orange line represents the $\epsilon=0$ isoline capturing the average activity $\mean{a_q}$ for a given current $q$. The $\lambda=0$ isoline exists only in the negative current half-plane, as it is there where the typical behavior occurs in the absence of bias on the current, while the $\epsilon=0$ isoline propagates through both the positive and negative currents half-planes since a given typical activity can be associated with both positive or negative currents. 

Using the measured joint LDF $G(q,a)$, see Fig.~\ref{fig:gqa}, and the univariate LDFs $F(q)$ and $I(a)$ obtained in Section \S\ref{s5}, see Fig.~\ref{fig:mG}, it is now possible to study the conditional LDFs $G_Q(q|a)$ and $G_A(a|q)$ defined in Eq. \eqref{eq:ldfcond}. This is shown in the right panels of Fig.~\ref{fig:gqa}. In particular, the top-right panel shows the current LDF conditioned to a fixed value of the activity, $G_Q(q|a)=G(q,a)-I(a)$, while the bottom-right panel displays the activity LDF conditioned to a fixed value of the current, $G_A(a|q)=G(q,a)-F(q)$. The LDF $G_Q(q|a)$ exhibits for $a>a_c$ two symmetrical current intervals around $q=0$ where it is affine or nonconvex, while for $a<a_c$ it is only defined for $q=0$, as expected from the joint activity-current fluctuation behavior observed in $G(q,a)$, see left panel in Fig.~\ref{fig:gqa}. Interestingly, biasing the activity to high values beyond its average behavior leads to an increase in the probability of negative current fluctuations. On the other hand, the probability of high positive current fluctuations is very small and almost independent of the activity $a$, see the tails of $G_Q(q|a)$ in the top-right panel of Fig.~\ref{fig:gqa}. The statistics of the activity conditioned on a given current is shown in the bottom-right panel of Fig.~\ref{fig:gqa}. The grey area in this plot represents the current intervals around $q=0$ where the bivariate $G(q,a)$ is affine or nonconvex. As $q$ increases, the mean activity conditioned on this value of the current grows, as does the probability of high activity fluctuations. Note also that the activity constraint on the current, $-a\le q \le a$, implies that $G_A(a<q|q)\to -\infty$ so $G_A(a|q)$ jumps discontinuously from a finite value to $-\infty$ at $a=|q|$. Finally, due to the time-symmetric character of the activity, the statistics of activity conditioned on a current $q$ is the same when conditioned on a current $-q$, and hence $G_A(a|q)=G_A(a|-q)$. This is another instance of the Gallavotti-Cohen theorem based on microscopic time-reversibility.

\begin{figure}
\bc
\includegraphics[width=13cm]{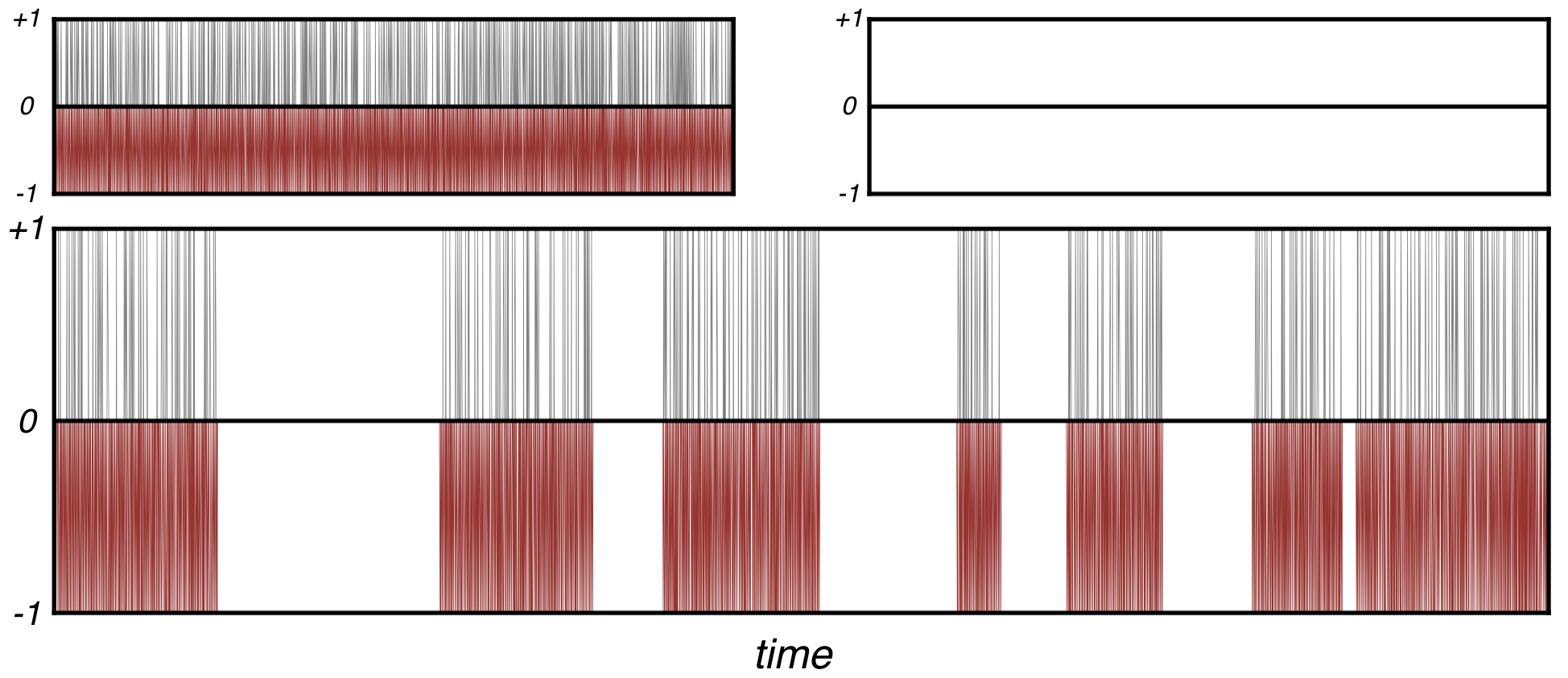}
\ec
\caption{Time evolution of dynamical parameters $C^+$ (grey) and $C^-$ (red) along representative quantum jump trajectories. Top left: No-dephasing case ($\gamma=0$) with a symmetric initial condition. Top right: No-dephasing case ($\gamma=0$) with an antisymmetric initial condition. Bottom: Small dephasing case ($\gamma=0.01$). In all panels the parameters are $B_z=0.5, \; \Gamma=0.1, \; \gamma= 0$ and $n=0.1$.}
\label{fig:jumps2}
\end{figure}

Even if these results are specific for a three-qubit system, we emphasize that similar features will arise for bigger systems. Interestingly, it has been shown that systems with more than three qubits can exhibit multiple strong symmetries and associated dynamical phase transitions \cite{manzano14a,manzano18a,thingna21}, as well as \emph{dynamical} symmetries \cite{buca21}. In this more complex scenario the system of interest would present several invariant subspaces, and the relation between current and activity might be more complicated than in the three-qubit case, though most of the phenomenology presented here may still hold.

To end this section, we now characterize the distinct dynamical patterns in the different symmetry phases of the dynamics, and how they coexist dynamically, with a distinctive intermittent pattern, in the presence of a weak dephasing noise channel. For that, we perform quantum Monte Carlo simulations of individual quantum jump trajectories, as in Section \S\ref{s3}, and measure an appropriate order parameter capable of distinguishing between both types of dynamics. As discussed in previous sections, in the antisymmetric state the system is reduced to a single qubit (as qubits 1 and 2 fall into a dark state and dynamically decouple) and the current is exactly zero while there is a net activity in the system. This is only possible because there is always one excitation entering and one coming out, a sort of excitonic blinking pattern. This type of locked blinking dynamics does not happen in general in the symmetric sector. To capture this essential dynamical signature we now define two different observables, $C^\pm$, such that $C^+=+1$ whenever \emph{two consecutive} quantum jumps introduce excitations in the system ($C^+=0$ otherwise), while $C^-=-1$ whenever two \emph{consecutive} quantum jumps remove excitations from the system ($C^-=0$ otherwise). Fig.~\ref{fig:jumps2} shows the time evolution of these two observables for different situations. In particular, the top panels show the dynamics of the three-qubits system in the absence of dephasing channel (i.e. when $\gamma=0$, see Eq. \eqref{eq:deph} and Section \S\ref{s2}), starting with a symmetric initial state (left) or an antisymmetric one (right). Clearly, dynamics in the symmetric case is characterized by many consecutive double jumps both up and down, but with net prevalence of double exciton removal jumps that leads to a negative average current in the symmetric steady state. On the other hand, the locked blinking dynamics in the antisymmetric sector implies that both $C^\pm$ remain exactly 0 along the whole time evolution, see top-right panel in Fig.~\ref{fig:jumps2}, and the current is always zero. This clear difference in the dynamics of $C^\pm$ confirms the validity of these observables as dynamical order parameters for the different symmetry sectors. The presence of dephasing noise, on the other hand, allows the mixing between both symmetry subspaces. Bottom panel in Fig.~\ref{fig:jumps2} shows the time evolution of our dynamical order parameters $C^\pm$ for a weak dephasing amplitude $\gamma=0.01$. Interestingly, the system evolution in this case exhibits \emph{intermittent behavior}, with periods of time where the system is trapped in the symmetric manifold, interrupted by jumps to the antisymmetric manifold allowed by the weak mixing introduced by the dephasing channel. This intermittent evolution is typical of a dynamical coexistence between phases of distinct activity, and is a direct consequence and a dynamical signature of the underlying symmetry of the three-qubits system. Such dynamical signatures can be harnessed to infer molecular symmetries from nonequilibrium transport experiments \cite{thingna16a}.

\section{Conclusions}
\label{s8}

In this paper we have investigated how strong symmetries affect both the transport properties and the activity patterns of a particular class of Markovian open quantum system, a three-qubits model under the action of a magnetic field and in contact with a thermal bath. Strong symmetries in open quantum systems lead to broken ergodicity and the emergence of multiple degenerate steady states \cite{buca12a}. A first observation in this work is that, interestingly, for initial states overlapping with several symmetry subspaces, individual quantum jump trajectories select randomly one of the symmetry sectors, collapsing in a finite time to the corresponding subspace and remaining there from that time on. This a particular instance of the dissipative freezing phenomenon recently observed in \cite{sanchez-munoz19a}, which implies a breakdown of a conservation law (associated to the underlying symmetry) at the individual trajectory level \cite{sanchez-munoz19a}. 

From a quantum jump perspective, the appearance of multiple steady states mentioned above is related to underlying dynamical phase transitions (DPTs) at the fluctuating level, that lead to a dynamical coexistence of different transport/activity channels classified by symmetry. We have studied these DPTs in the univariate large deviation functions associated to the exciton current (a key time-antisymmetric observable characterizing transport out of equilibrium) and the dynamical activity (a time-symmetric magnitude of direct experimental relevance which may constraint the range of current fluctuations). In particular, we find a pair of twin dynamical phase transitions in exciton current statistics, induced by the strong symmetry and related by time reversibility, where a zero-current antisymmetric phase (under the exchange of qubits 1 and 2) coexists with a symmetric phase of negative exciton current. On the other hand, the activity statistics exhibits a single DPT (since the activity is a time-symmetric observable) where the symmetric and antisymmetric phases of different but nonzero activities dynamically coexists. Interestingly, the maximum current and maximum activity subspaces do not coincide for the three-qubits model studied here, as the maximum current subspace corresponds to the antisymmetric one while the maximum activity subspace corresponds to the symmetric sector.

In addition, this work also discusses how symmetries are reflected in the \emph{joint} large deviation statistics of the activity and the current, a central observable in order to fully characterize the complex, coupled quantum jump dynamics both in the time-symmetric and time-antisymmetric sectors. The presence of a strong symmetry under nonequilibrium conditions implies non-analyticities in the dynamical free energy in the dual activity-current plane (or equivalently in the joint activity-current large deviation function). Remarkably, the DPT predicted around the steady state and its Gallavotti-Cohen twin dual are extended into lines of first order DPTs in the current-activity plane, with a nontrivial structure which depends on the transport and activity properties of each of the symmetry phases (in particular on the average current and activity of each of these phases). We further find that activity constraints the range of current fluctuations, leading in particular to an activity-driven current lockdown phase for activities below some critical threshold, a new route to control quantum transport enabled by symmetry. The presence of a noisy dephasing channel acting on all qubits breaks the exchange symmetry of the three-qubits system, restoring global ergodicity and leading to an unique steady state. This dephasing noise also washes out the symmetry-induced DPTs, although the underlying topological symmetry leaves a dynamical fingerprint in the form of an intermittent, bursty on/off dynamics between the different symmetry sectors when the dephasing amplitude is weak, a phenomenon observed in quantum Monte Carlo simulations of individual quantum jump trajectories, and well-captured by some blinking order parameters $C^\pm$.

\section*{Akcnowledgements}
We acknowledge the Spanish Ministry and \emph{Agencia Estatal de Investigaci{\'o}n} (AEI) through grant FIS2017-84256-P (European Regional Development Fund), as well as the \emph{Consejer\'{\i}a de Conocimiento, Investigaci\'on y Universidad, Junta de Andaluc\'{\i}a} and European Regional Development Fund, Ref. A-FQM-175-UGR18 and SOMM17/6105/UGR for financial support. We are also grateful for the computational resources and assistance provided by PROTEUS, the supercomputing center of Institute Carlos I for Theoretical and Computational Physics at the University of Granada, Spain.

\section*{Bibliography}
\bibliographystyle{unsrt}


\end{document}